\documentclass[referee,useAMS,usenatbib,usegraphicx]{mn2e}

\newcommand{\aap}{A\&A}
\newcommand{\aj}{AJ}
\newcommand{\apj}{ApJ}

\newcommand{\apjs}{ApJS}
\newcommand{\araa}{ARA\&A}
\newcommand{\mnras}{\mbox{MNRAS}}

\title[Large-Scale Gas Dynamics ]{Large-Scale Gas Dynamics
in the Adhesion Model: Implications for the Two-Phase Massive Galaxy Formation Scenario}
\author[Dom\'{\i}nguez-Tenreiro et al.]{R. Dom\'{\i}nguez-Tenreiro$^{1}$\thanks{E-mail:
rosa.dominguez@uam.es}, J. O\~norbe$^{1}$, F. Mart\'{\i}nez-Serrano$^{2}$ and A. Serna$^{2}$\\
$^{1}$Departamento de F\'{\i}sica Te\'orica, C-XI. Universidad Aut\'onoma de Madrid, Madrid, E-28049, Spain\\
$^{2}$Departamento de F\'{\i}sica y A.C., Universidad Miguel Hern\'andez, Elche,
Spain}
\begin{document}

\date{Accepted . Received ; in original form }

\pagerange{\pageref{firstpage}--\pageref{lastpage}} \pubyear{}

\maketitle

\label{firstpage}

\begin{abstract}

We have studied the mass assembly and star formation histories of massive galaxies  
identified at low redshift $z$ in different cosmological hydrodynamical simulations.
To this end, we have carried out
a detailed follow-up backwards in time
of their constituent  mass elements (sampled by particles)
 of different types.  After that,  the configurations they
depict at progressively higher $z$s  were carefully analysed.

The analyses show that these histories share common generic patterns,
irrespective of particular circumstances.
In any case, however, the results we have found are different depending on the particle type.
The most outstanding differences follow.

We have found that by $z \sim 3.5 - 6$, mass elements
identified as stellar particles at $z = 0$ exhibit a gaseous cosmic-web-like
morphology with scales of $\sim 1$ physical Mpc, where the densest mass
elements have already turned into stars by $z \sim 6$. These settings are in fact
the densest pieces of the cosmic web, where no hot particles
show up, and dynamically organised as a  hierarchy of
{\it flow convergence regions} (FCRs), that is, attraction basins for mass
flows.  
At high $z$ FCRs undergo fast contractive deformations  with very low angular momentum,  violently shrinking them. Indeed, by $z \sim 1$
most of the gaseous or stellar mass they contain shows up as
bound to a massive elliptical-like object  at their centers, with typical
half mass radii of $r_{\rm star}^{\rm mass} \sim 2 - 3$   kpc.    
After this, a second phase comes about where the mass assembly rate
is much slower and
 characterised by mergers involving angular momentum.

On the other hand, mass elements identified at the diffuse hot coronae
surrounding  massive galaxies at  $z = 0$  
do not display a clear web-like morphology
at any $z$. Diffuse gas is heated when FCRs
go through contractive deformations. Most of this gas  remains hot
and with low density  throughout the evolution.

To shed light on the physical foundations of the behaviour  revealed by our analyses 
(i.e., a two-phase formation  process with different implications for diffuse or shocked mass elements),
as well as on their possible observational implications,
these patterns have been confronted with  some generic properties
of singular flows as described by the adhesion model
(i.e., potential character of the velocity field, singular versus regular points,
dressing, locality when an spectrum of perturbations is implemented).
We have found that  the common patterns the simulations show
 can be interpreted as a  natural consequence of  flow properties  
that, moreover, could explain different generic observational results on
massive galaxies or their samples. We briefly discuss some of them.

\end{abstract}


\section{Introduction}
\label{Intro}

Among all galaxy families, ellipticals\footnote{ Low-redshift galaxies divide into two distinct families at a stellar mass of $ \approx 3 \times 10^{10} $ M$_{\odot}$ \citep{Kauffmann:2003a,Kauffmann:2003b}, with more massive galaxies having the characteristics
of elliptical galaxies. In this paper the term  {\it elliptical}  will most often be used when we refer to observations.}
 are  the simplest
and those that show the most precise regularities,
sometimes in  the form of power-law correlations
between  some pairs of their observable parameters.
The Sloan Digital Sky Survey \citep[SDSS]{York:2000}
has substantially improved the statistics on elliptical, hereafter E,
samples.
Analyses of the distributions of their luminosities $L$,
projected effective radii $R_e$ and
central line-of-sight  velocity dispersions $\sigma_{\rm los, 0}$
 \citep{Hyde:2009},
indicate that they follow Fundamental Plane relations, among other correlations.
These correlations are consistent with those previously
established in the literature \citep[see references and discussion
in][]{Bernardi:2003b,Bernardi:2003c,Bernardi:2003d}
and are thought to
carry fundamental physical information on the processes
involved in the assembly of Es.

Recently, spectral indices have been identified
(H$_{\beta}$, H$_{\gamma}$, H$_{\delta}$) that break
the age-metallicity degeneracy,
allowing for an improved  stellar age determination
in E galaxies through  evolutionary synthesis models
\citep[see review in][]{Maraston:2003,Sanchezblazquez:2006}.
Even if still hampered by uncertainties, these
models point now to more massive Es having older mean ages
and lower rates of recent star formation
and, also,
higher suprasolar $\alpha/<$Fe$>$ ratios
than less massive ones.
The values of these ratios and their
correlation with mass indicators
suggest that an important fraction of the stars
in most massive galaxies
formed on short time-scales, and that this
fraction increases with $\sigma_{\rm los, 0}$ or galaxy mass.
\citep[see][and references therein]{Thomas:2005,Gallazzi:2006,Jimenez:2007,Clemens:2009}.
These age effects link elliptical dynamical properties with the
characteristics of their stellar populations, and are another manifestation of the physical
regularities underlying massive    galaxies.
Such an effect is also known as the
 downsizing phenomenon, first introduced by \citet{Cowie:1996}.
Understanding the origin of such regularities in massive galaxy samples is a very important task
and it is now affordable.

Two  views have historically existed on how elliptical galaxies
formed. In the classical {\it monolithic collapse scenario}
\citep{Eggen:1962,Tinsley:1972,Larson:1974},
 ellipticals form at high $z$ in a single
burst of star formation ensuing the collapse of a gas cloud.
The modern version of the  monolithic collapse scenario
puts the stress on the assembly out of gaseous material (that is,
with dissipation). This assembly can be
either in the form of  a unique cloud or
of  many gaseous clumps,   but not from 
pre-existing stars. Indeed,
the stellar populations formed
at high $z$ and on short time-scales
relative to spirals \citep{Matteucci:2003}.
The competing {\it  hierarchical scenario} \citep[e.g.][]{Toomre:1977,White:1978}
propounds that galaxies form hierarchically  through successive random
mergers of subunits (the so-called galaxy merger tree)
over a wide redshift range,
in such a way  that  more massive ones (that is, ellipticals)
are more likely to form  at  later times.
The stress here is on no dissipative assembly
through random mergers and on
large formation time-scales for the stellar populations.

The regularity
of the E family shown by the correlations described above,
as well as age effects in massive Es,
seem to favour the monolithic collapse scenario.
In fact, they are difficult to explain in models where
massive galaxies are assembled at late times by random  mergers
of pre-existing subunits,  as in
the {\it standard hierarchical model} of galaxy
formation.
The monolithic scenario also explains another set of observational results on
E galaxy homogeneity, such as, for example,
i), the lack of significant structural and
dynamical evolution of lens E galaxies, at least out to $z \sim 1$
\citep{Treu:2004}; ii), the lack of any strong structural evolution in the
Fundamental Plane  relation since $z \sim 3$ \citep{Trujillo:2004,McIntosh:2005};
 and iii), the confirmed existence of a population of  old,
relaxed, massive ($M^{\rm star} > 10^{11}$M$_{\odot}$) spheroidal galaxies at
intermediate $z$s \citep[$z \sim 1 - 2$,][]{Cimatti:2002,Cimatti:2004,Stanford:2004},
or that of massive objects with old stellar populations  earlier on at $z \sim 4-5$
 \citep{Mobasher:2005,Wiklind:07,Mobasher:2010}.

However, the monolithic scenario
does not recover all the currently available observations  on Es,
either. Important examples are:
i), the growth of the total stellar mass bound up in bright red galaxies
by a factor of $\sim 2$ since $z=1$
\citep{Bell:2004,Conselice:2005,Drory:2004,Fontana:2004,Bundy:2005,Faber:2007}, implying that
the mass assembly of most Es continues  below $z = 1$;
ii),  the signatures of merging 
\citep{LeFreve:2000,Patton:2002,Kartaltepe:2007,Lin:2008,Lotz:2008,Conselice:2009,Bridge:2010}, 
in particular of major dissipationless mergers between massive
galaxies \citep{Bell:2005,Bundy:2009},
that translate into a relatively high  merger rate  even  below $z = 1$; and
iii), the need for a young stellar component in some E galaxies
\citep{vanDokkum:2003,vanderWel:2004},
or, more particularly,
the finding of blue cores, that is, recent  star formation
at the central regions \citep[see][and references therein]{Menanteau:2004}.
 These examples suggest that mergers at $z$s below
$\sim 1.5 - 2$ could have played an important role in massive galaxy assembly.

%%%%%%%%%%%%%%

Other sets of observational data that any scenario on massive galaxy formation
has to interpret include:
i), the existence of a diffuse, X-ray emitting gaseous corona around
ellipticals, and the correlations their properties show with
those of the galaxy they surround   \citep{Beuing:1999,Diehl:2005};
ii), the correlations among black hole properties at the  centres
of ellipticals and those of their hosting galaxy
\citep{Magorrian:1998,Gebhardt:2000,Ferrarese:2000,Ferrarese:2005}, as well as the starburst-AGN
connection \citep{Aretx:2001}; and
iii), the relative stability of massive galaxies shapes at low
redshifts, while less massive galaxies  acquire on average their
stable shapes later on \citep{Zheng:2005,Neichel:2008}.

In order to reconcile all this observational background
within a formation scenario, it is preferable to study
galaxy assembly from simple physical principles and in
connection with the global cosmological model.
Self-consistent gravo-hydrodynamical simulations are a very
convenient tool  to work this problem out
\citep{Navarro:1994,Tissera:1997,Thacker:2000}.
Individual galaxy-like objects including massive ones  naturally appear as an output
of the simulations, and no  prescriptions  are needed
as far as their mass  assembly processes
are concerned at scales of a few  hundred kpc.

On the other hand, the results of a self-consistent
simulation  will be more easily
understood in the context of   physical theories for the
advanced non-linear stages of gravitational
instability,
just as the results of an  experiment can be better understood
or interpreted when experimenters  have a theoretical background
at their disposal.
Such a physical theory is provided by the Zeldovich approximation \citep{Zeldovich:1970}
and its extension
to the adhesion model \citep{Gurbatov:1984,
Gurbatov:1989,Shandarin:1989,
Gurbatov:1991,Vergassola:1994},
including singularity dressing \citep{Dominguez:2000} and gas physics.

The adhesion approximation has already been used in the context
of N-body simulations to predict when and where  large scale singularities
or caustics (i.e., the skeleton of the large scale mass distribution)
form  \citep{Kofman:1990,Weinberg:1990}. In this case,  it showed
its potentialities to study a wide class of cosmological problems
at these scales. Here we will use it as a theoretical framework
to try to understand the mass assembly of massive galaxies  as an accretion process onto caustics
or "caustic dressing" at smaller scales, including gas processes.
Indeed, the aim of this paper is to study hydrodynamic simulations in more detail
in this context.
More particularly, we analyse the implications that different
aspects of the dynamics of singular flows in an expanding universe
(singularity patterns, dressing, locality)
could have on the mass assembly processes of massive galaxies.
In other words, we aim at  deepening the links among some aspects
of flow dynamics and different observational characteristics
of massive galaxies recently discovered, and that could be a
consequence of how these galaxies have acquired their baryons at
scales of some few hundred kpc.
To avoid that these links get hidden by detailed subresolution
modelling, we will  remain at  the simplest level in this respect.

This paper is organised as follows.
The theoretical context  is briefly introduced in
$\S$\ref{Hypo}.
The simulation method is presented in $\S$\ref{Methods},
where we also very briefly outline
the comparison of the properties of massive objects formed
in the simulations with observational data on ellipticals.
In $\S$\ref{TestHypo} we analyse cosmological hydrodynamical simulations.
In $\S$\ref{implications} we summarise the main results of this paper,
we discuss mass assembly within the theoretical background
put forward in  $\S$\ref{Hypo} as well as  some of its
generic observable implications, and, finally,
we give the conclusions of this work.

\section{SINGULARITY PATTERNS, DRESSING, LOCALITY}
\label{Hypo}

The advanced non-linear stages of gravitational instability
are described by
the {\it adhesion model} (Gurbatov \& Saichev 1984;
Gurbatov, Saichev \& Shandarin 1989; Shandarin \& Zeldovich 1989;
 Gurbatov et al. 1991; Vergassola et al. 1994),
 an extension  of Zeldovich's
(1970) popular non-linear approximation.
In comoving  coordinates, Zeldovich's approximation  is:
\begin{equation}
x_i(\textit{\bf q},b(t)) = q_i + b(t) v_i(\textit{\bf q})
\end{equation}
where $q_i$ and $x_i,  i = 1,2,3$ are comoving  Lagrangian and
Eulerian
coordinates of fluid elements or particles sampling them
 (i.e., initial positions at time $t_{in}$ and positions at
later  times $t$),
 respectively; $b(t)$ is the function
of time describing the evolution of the  growing density mode
in linear gravitational instability and taken to be the time variable
in the Zeldovich approximation; and
$v_i(\textit{\bf q}) \equiv  V_i(\textit{\bf q})/ \dot{b} a $,
 with $V_i(\textit{\bf q})$  the
initial peculiar  velocity field and $a(t)$ the cosmic scale function.

As it is well known,
Zeldovich's solution is not applicable beyond particle crossing,
because it predicts that caustics thicken and vanish due to
multistreaming soon after their formation.
However, N-body simulations of large-scale structure formation indicate
that long-lasting pancakes are indeed formed, near which particles stick:
multistreaming did not take place.
The adhesion model was
introduced to incorporate this feature to
Zeldovich's approximation.

One way of avoiding multistreaming is to introduce a small diffusion
term in Zeldovich's momentum equation, in such a way that it has an
effect only when and where particle crossings are about to take place.
This can be accomplished by introducing a non-zero viscosity,
$\nu$, and  then taking the limit $\nu \rightarrow 0$.
This is  the phenomenological derivation of the  adhesion model. 
 Physically motivated derivations
can be found in \citet{Buchert:1998}, \citet{Buchert:1999} and others included in
the review by  \citet{Buchert:2005}.
As in the Zeldovich approximation, in the adhesion model
 motion is potential. Hence,
the initial velocity field can be expressed as the gradient
of a scalar potential field, $\Phi_0(\textit{\bf q})$,
describing the spatial structure of the initial perturbation field.
It can be shown that  the solutions
for the velocity field behave just as those of Burgers' equation
\citep{Burgers:1948,Burgers:1974} in the limit $\nu \rightarrow 0$,
 whose analytical solutions
are known.

The most significant characteristic of Burgers' equation solutions
  is that they are discontinuous and   hence they unavoidably develop shocks,
i.e., locations where
at a given time the velocity field becomes
discontinuous and certain particles coalesce  into long-lasting
singularities with different geometries\footnote{
Shocks are also called caustics and both names will be used henceforth. 
Shocks should not be mistaken for shock waves, that is,
discontinuities that travel inside fluids heating the subvolume
in their path and causing its  entropy to increase.
}.
Mathematically,  caustics at time $b(t)$
 can be considered as   singularities in
the so-called
{\it Lagrangian map} that transforms the initial point configuration
(i.e.,  Lagrangian coordinates, $\textit{\bf q}$) into the point configuration
at  time $b(t)$ (Eulerian coordinates, $\textit{\bf x}(\textit{\bf q},b)$).
A singularity occurs at time $b(t)$
when a non-zero $d$-dimensional volume $V$ around point $\textit{\bf q}$
 in the initial point configuration is
mapped to a $d'$-dimensional (with $d'$ lower than $d$) volume
around point   $\textit{\bf x}(\textit{\bf q},b)$ in Eulerian space.
The mass involved in a given caustic is
proportional to the volume $V$.
In a three dimensional
space ($d$=3), depending on their dimensionality,
 caustics can be walls ($d'$=2 or surfaces in the Eulerian $\textit{\bf x}$ space),
filaments ($d'$=1),
 and nodes ($d'$=0).

The adhesion model implies that walls are first formed as denser
small surfaces (the so-called pancakes), then they grow until
they intersect each other along filaments, that on their turn
intersect at nodes. This completes the formation of the cellular structure at
a given scale.
A further complication is that walls, as two-dimensional systems,
usually develop filamentary
singularities whose formation follows the same patterns as wall formation
in the three-dimensional space, but now with $d=2$ and $d'=1$.
By the same reason, filaments (either  in the full 3D cell
structure or within walls) usually fragment and develop nodes
where a fraction of their mass ends up. We will call them secondary filaments
or nodes to distinguish them from primary filaments (at wall intersection)
or primary nodes (at primary filament intersection).
Secondary filaments or nodes can be seen in numerical simulations. However, they
are not predicted by the adhesion model in three dimensions, but are recovered in two
and one-dimensional models, respectively.

At nodes, and more particularly at primary ones,
  mass piles up.
So, at a given scale, walls, filaments and nodes (the cosmic web) are successively
formed. Walls and filaments are the paths of shocked particles
towards nodes and, at a given scale, they are not long-lasting
configurations, but  rather vanish as the mass piles up at nodes.
This is how the cell structure is erased at given scales.
It can be shown  \citep{Vergassola:1994} that this behaviour is determined by
the structure of
the minima of the potential function $-\Phi_0(\textit{\bf q})$ and that
asymptotically (i.e., at large $b$) the behaviour of the system
 is controlled only by its {\it deepest minima}.
Indeed,
cells  swallow up  some of their
neighbouring cells  associated with less deep minima
of $-\Phi_0(\textit{\bf q})$, involving their constituent elements
(i.e., walls, filaments and nodes).
This causes contractive flow deformations that erase
substructure at cell scales through the coalescence of these elements
and their mass piling up into essentially a unique node
(i.e.,  a kind of collapse event).
We see that the advanced stages of non-linear gravitational evolution act as a
kind of short-scale smoothing process on the cosmic web at scales increasingly larger,
while the web is still forming at even larger scales.
%initial density fluctuations.
As time elapses, the structure evolution becomes dominated by larger and larger modes of the
density field and finer details are removed by merging and collapse of
these modes.

Note that this mass piling up  occurs mathematically at {\it zero
relative angular momentum}
 due to the potential character of the initial
velocity field.
Moreover, and very significantly for massive galaxy formation,
Burgers' equation solutions ensure the existence
at any time $b(t)$ of {\it regular points or mass elements},
as those that have not yet been trapped into a caustic
at $b(t)$.
Because of that, these regular mass elements
are among the  least dense in the density distribution.
Note, however, that due to the complex structure of the flow,
singular (i.e., already trapped into a caustic) and
regular (i.e., not yet trapped) mass elements need not be spatially segregated,
and in fact, they are mixed ideally at any scale.

One could think
that singular mass elements would end up mostly
within zero volume singularities.
Remember, however, that the phenomenological adhesion model
tells nothing about the internal density or velocity structure of
locations where mass gets adhered.
Otherwise, numerical simulations indicate	that most
mass  reaching nodes ends up in virialised structures,
 where gravitational
attraction is  balanced by velocity dispersion.
How does this come about?
Just to have a clue from theory, we recall that
in his derivation of a generalised adhesion-like model,
\citet{Dominguez:2000} finds  corrections to the momentum
equation of the Zeldovich approximation that regularises
(i.e., dresses) its singularities.  These then become long-lasting structures where more mass gets stuck,
but within non-zero volumes supported by dispersion
\citep[see also][for a discussion of these effects
in terms of the viscosity phenomenologically introduced in
the adhesion model]{Gurbatov:1989}.
The analyses of N-body simulations strongly suggest that
any kind of flow singularity gets dressed \citep[i.e., not only at pancakes,
as it has been analytically proven by][]{Dominguez:2000}.  Therefore, flow singularities become visible as the places where mass piles up.

 Indeed, it is known that in N-body simulations, particles are stopped
at and get spatially  confined in the
neighbourhood of nodes, as  they change their macroscopic energy into
velocity dispersion. In this way,
they give rise to self-gravitating configurations in
virial equilibrium (i.e., massive halos).
These configurations are
 characterised by  mass, velocity dispersion
 and size scales $M_{\rm vir}, \sqrt{< v^2 >}$ and
  $r_{\rm vir}$, respectively, linked by the virial theorem
\citep[involving also a shape factor of order unity, see][]{Binney:2008}.
We note that this {\it purely mechanical} dissipative mechanism,
 acting on  any kind of gravitating matter, has
the same origin  as the  viscous-like forces in a gas.
It has also the same consequences, except that pure
gravitating matter cannot lose its energy  through radiation or
other cooling mechanisms.
When gas is added, the energy transfer from ordered
to disordered motions  includes
the transformation of velocity dispersion
into internal energy (heating) and pressure,  and then  
energy is lost through cooling, mainly at the densest pieces of the cosmic web.
The consequences of these processes cannot be deciphered from theory alone but  additionally need gravo-hydrodynamical
simulations.

In Cosmology one usually assumes that the field of initial density perturbations is
gaussian and with power spectrum given by
$P(k) \equiv \langle  \mid \delta_k \mid ^2 \rangle = A k^n$,
where $\mid  . \mid $ means module and  brackets
mean statistical  average.
In this case, the statistical properties of Burgers' equation
solution  in the $\nu \rightarrow 0$ limit are
scale invariant in space and time ($b$) variables, see details in \citet{Vergassola:1994}.
As a consequence, a length-scale appears in the system, the coalescence
length defined by $L_c(t) = (Cb(t))^{2/(n+3)}$ (here $C$ is a normalisation constant such that
$\langle (\delta v_l)^2 \rangle = C l^{-(n+1)}$,
where $\delta v_l \equiv \langle  \mid \vec v(\vec x + \vec l) - \vec v(\vec x) \mid \rangle $).
Its physical meaning is as follows:
at a given time, $ b(t)$,
coalescence dominates and substructure is substantially erased
at scales $l < L_c(t)$, while the initial situation is expected
to have changed only very marginally at larger
scales.
It is worth noting that $L_c(t)$ is defined based on  global average
values. Note, however, that those regions $R$ where typically
$ (\delta v_l)^2 > \langle (\delta v_l)^2 \rangle $
 have {\it local}
coalescence lengths at a given time larger than average, $L_c(R, t) > L_c(t)$
(and conversely). Therefore, in these regions evolution proceeds faster
(more slowly) than average.
 Also, scales of a given size $l$ suffer a contractive
deformation at different
times depending on where they are  placed, and they involve different
masses.
As we will see below, this {\it locality} has important consequences
to explain some massive galaxy properties.

We now turn to more specific issues related to
mass assembly in numerical simulations.
In simulations using particles  to sample fluid elements, it is very easy to
know where the particles that constitute a given
object at a low $z = z_{low}$ are initially at $b(t_{in})$.
In fact, mathematically this represents the image by the {\it inverse
Lagrangian map} of the set of constituent particles at $z_{low}$.
We will call the
configuration they define  in the Lagrangian $\textit{\bf q}$ space
the  {\it proto-object region}
for this object (hereafter, POR). The POR of an object can be visualised
through the positions at $b(t_{in})$ of its  particles.
When we consider an arbitrary $z$, we will talk about
the POR of the object {\it at $z$} instead,  hereafter  $PO(z)$.  Again, it
can be visualised through the
positions at this redshift $z$ of the constituent particles of the
object (see $\S$\ref{TestHypo}).

We now introduce a useful concept to describe the
mass assembly at high $z$  of an   object that is bound at a lower $z_{low}$:
 the {\it caustic or shock tree}
for this object.
Consider the set of  constituent or bound particles of the object at
$z_{low}$. The object caustic tree is defined as the set  of
 caustic formation events involving  these particles
 (walls, filaments, nodes as well as their
fusions at different
scales and at different times) from $b(t_{in})$
 until they become bound to the assembled object.
  It is a generalisation  of the merger tree concept
in the Press-Schechter theory, enlarged to take into
consideration the richer variety of possibilities here.
It represents the development of the PORs of the object at different $z$s.
The object gets assembled through its caustic tree
involving the particles that at $b(t_{in})$ constitute its POR, whose
transformations by the Lagrangian map/caustic tree can be visualised
through the corresponding $PO(z)$s. We remind that, as noted above,
the caustic tree development  of a given object is already contained in the
 structure of  the minima of $-\Phi_0(\textit{\bf q})$  at the object POR.

The description of gravitational instability we have outlined,
based on the adhesive behaviour of matter in a
fluctuating density field, can be recasted
in the language of gravitational collapse \citep{Padma:1993}
and the Press-Schechter theory \citep{Press:1974}.
In particular, $L_c(t)$ is related to the typical mass
$M_c(t)$
of collapsed objects (turnaround) at time $t$
appearing in the theory of non-linear spherical
collapse. The main difference lies in the geometry (not necessarily
spherical or even three dimensional) of the mass accumulation
process, and in the scale invariance properties of the
solutions  summarised in the coalescence length;
these are useful concepts to interpret some massive galaxy properties.
Turnaround is seen here as a node formation,
that is, in the language of the Lagrangian map, as  a
  {\it contractive   deformation} acting on
Lagrangian volumes (see above),
and, as already said, the merger tree is now the caustic tree.
However, the distinction is important, because of its phenomenological
implications on galaxy  formation \citep[see for example][]{Jones:2010, Aragon:2010},
and more particularly in mass assembly at high $z$, as we will see.

\section{Methods}
\label{Methods}

The quantitative aspects of mass assembly in massive galaxies can only be
known through gravo-hydrodynamical simulations, conveniently analysed
and visualised.

\subsection{Codes and SF Implementation}
\label{CodeSF}

The simulations analysed here are mainly based on the GALFOBS
project\footnote{See at http://www.deisa.eu/science/deci/projects2007-2008//GALFOBS}.

We have used
the \texttt{P-DEVA} code. This is an  OpenMP AP3M-SPH code specially designed to study galaxy assembly in
 a cosmological context \citep{TesisFran}.
In this code, as well as in its previous sequential version DEVA,
particular attention is paid that the conservation laws (energy, entropy,
momentum and angular momentum) hold as accurately as
possible\footnote{This in particular implies that a double loop in the
 neighbour searching algorithm must be used, which considerably
increases the CPU time} \citep[see][for details]{Serna:2003}.

Star formation processes have been included through a simple
parametrisation \citep{Katz:1992} that transforms   locally-collapsing gas
at kpc scales,
denser than a threshold density,
$\rho_{\rm thres} $,
into stars at a rate
$d\rho_{\rm star}/dt = c_{\ast}  \rho_{\rm gas}/ t_g$. Here
$t_g$ is a characteristic time-scale chosen
to be equal to the maximum of the local gas-dynamical time,
$t_{dyn} = (4\pi G\rho_{\rm gas})^{- 1/2}$,
and the local cooling time;   $c_{\ast}$ is the average
star formation efficiency at kpc scales.
This implementation of star formation is equivalent to
the Kennicutt-Schmidt law.
For details on the SF implementation, see
\citet{Onorbe:2007}.
No explicit SF feedback or other discrete energy injection
mechanisms have been considered. However, they have implicitly taken into account
through the particular values given to the $\rho_{\rm thres} $
and  $c_{\ast}$ parameters.
We have chosen to keep at this simple level of subgrid physics
modelling
because our aim in this paper is to test how far  we can reach relative
to massive galaxy formation just from the
generic properties of singular flows as introduced in $\S$\ref{Hypo}.

\subsection{Runs}

We report here
on results of hydrodynamical simulations run in the context of
a $\Lambda$CDM cosmological model whose
 parameters, as well as  those of
the field of primordial density fluctuations (i.e., initial spectrum),
have been taken from CMB anisotropy data \citep{Dunkley:2009}
with priors\footnote{http://lambda.gsfc.nasa.gov/product/map/current/params/lcdm\_sz\_lens\_run\_wmap5\_bao\_snall\_lyapost.cfm}.
The simulations have been implemented in a periodic box of
80 Mpc comoving  side, with a gravitational softening of $\epsilon_g = 2.3 $ kpc
and a minimum hydrodynamical smoothing length half this value.
 The mass of dark matter and baryonic particles are
1.5 $\times 10^{8}$ and 2.4 $\times 10^{7} $ M$_{\odot}$, respectively.
This gives 512$^3$ dark matter and 512$^3$ baryonic particles in the 80 Mpc side box
(main simulation), a size that attains cosmological convergence
\citep{Power:2006}.

When analysing galaxy formation in numerical  simulations, it is desirable to
be sure that the galaxies the simulation produces are consistent with
observations at low $z$s.
Due to the extreme CPU consumption by hydrodynamical forces, this is not yet possible
for the main GALFOBS simulation.
However, we have been able to reach $z=0$ for several sub-volumes of this main simulation.
To do so, we have run several simulations in which we have
just  computed the gravitational forces for the full box,  
and the hydrodynamical forces only in a certain region (sub-boxes of 26 Mpc  comoving side) out of the total volume. 
Care has been taken that these sub-volumes
sample different environments.
In order to not have problems with the interaction of the different sub-volumes,  for our analyses we used the data
well inside them, in this case  cubes of  20 Mpc  comoving side.
 Five such sub-boxes  reached
$z = 0$ or $z \approx 0$, (S2100, S2109, S2110, S2114 and S2115)
 and massive galaxy objects (hereafter MGOs)  in there  were analysed
and found to be consistent with observational data
of local ellipticals, see \citet{TesisJose}.

To compare the mass assembly and SF patterns of MGOs
formed in these simulations with those of MGOs
obtained in our previous simulations \citep{Onorbe:2006,Onorbe:2007},
several of those have been also studied,  focusing on S8743.
We recall that the DEVA code  was used to run these simulations. Also, 
the primordial spectrum of fluctuations  was implemented
in a periodic box of 10 Mpc comoving side  sampled by 64$^3$ dark matter
and 64$^3$ baryonic particles, with the same gravitational softening as
in the large box simulation.
The initial spectrum normalisation  was taken slightly  high
($\sigma_{8}$=1.18) to mimic an active  region of the universe.

\begin{table*}
\centering
\begin{minipage}{160mm}
\caption{Parameters of the simulations and some MGO parameters. (1): Simulation run; (2): $H_{0} / 100\,km\,s^{-1}$; (3): Matter density; (4): Baryon density; (5): Power spectrum normalisation; (6): Density threshold; (7): Star formation efficiency; (8) Number of dark matter and baryonic particles; (9) Smoothing length ($h^{-1}Mpc$); (10) Virial radius (kpc) (11) Central velocity dispersion ($km\times s^{-1}$) (12) Stellar effective radius (kpc) (13) Stellar mass ($10^{10}$ $M_{\odot}$).}
\label{tab:simparam}
\begin{footnotesize}
\begin{tabular}{@{}c cccc cc cc cccc@{}}
\hline
Simulation &  $h$ & $\Omega_{m}$ &  $\Omega_{b}$  & $\sigma_{8}$ &$\rho_{thres}$ & $c_{*}$ & $N_{DM}+N_{bar}$ & $\epsilon$ &  $r_{\rm vir}$ & $\sigma^{\rm star}_{\rm los,0}$ &  $R_{\rm star}^{\rm e, bo}$ & $M^{\rm star}_{\rm bo}$\\
(1) & (2) & (3) & (4) & (5) & (6) & (7) & (8) & (9) & (10) & (11) & (12) & (13) \\
\hline
S2100 & 0.694 & 0.295 & 0.0476 & 0.852  & $4.8\times10^{-25}$ & 0.3 & see text & 0.0015 & 450 & 233.64 & 4.47 & 29.86 \\
S8743 & 0.65 & 0.35 & 0.06 & 1.18 & $6\times10^{-25}$ & 0.3 & $64^{3}+64^{3}$  & 0.0015 & 532 &  217.03 & 6.38 & 26.40 \\
\hline
\end{tabular}
\end{footnotesize}
\end{minipage}
\end{table*}

\subsection{Results \& Comparisons to Observational Data}
\label{comparisons}

Galaxy-like objects of different morphologies
(disk-like, elliptical-like, irregulars) have
been identified in  any kinds of simulations.
The analyses of their structural and
dynamical properties show that they are consistent with observations of the
local universe, both for disk-like \citep{Saiz:2001,MartinezSerrano:2009} and elliptical-like
objects \citep{Saiz:2004,DTal:2004,Onorbe:2006,Onorbe:2007,GonzalezGarcia:2009}.
For details on the mass and velocity distributions of the different
components, and on the parameters characterising them,
we refer the reader to these publications.

\section{ THE PROCESS OF MASSIVE GALAXY ASSEMBLY}
\label{TestHypo}

These agreements with
 observations  indicate that the assembly patterns of
 simulated objects could mimic key aspects of real massive galaxy assembly
 patterns.
To deepen into the physics underlying such patterns in a
cosmological context, we have analysed in terms of the cosmic web dynamics
  the sequence of events giving rise to MGO formation in the  preceding simulations.
   Here we give a description of their common characteristics by
   focusing on two such objects
(hereafter MGO \#1 and MGO \#2), formed in two different
   simulations (S2100 and S8743),
whose  characteristics are described in Table~\ref{tab:simparam}.
Some resulting properties of the two MGOs are also given in this Table.

\begin{figure}
\includegraphics[width=.8\textwidth]{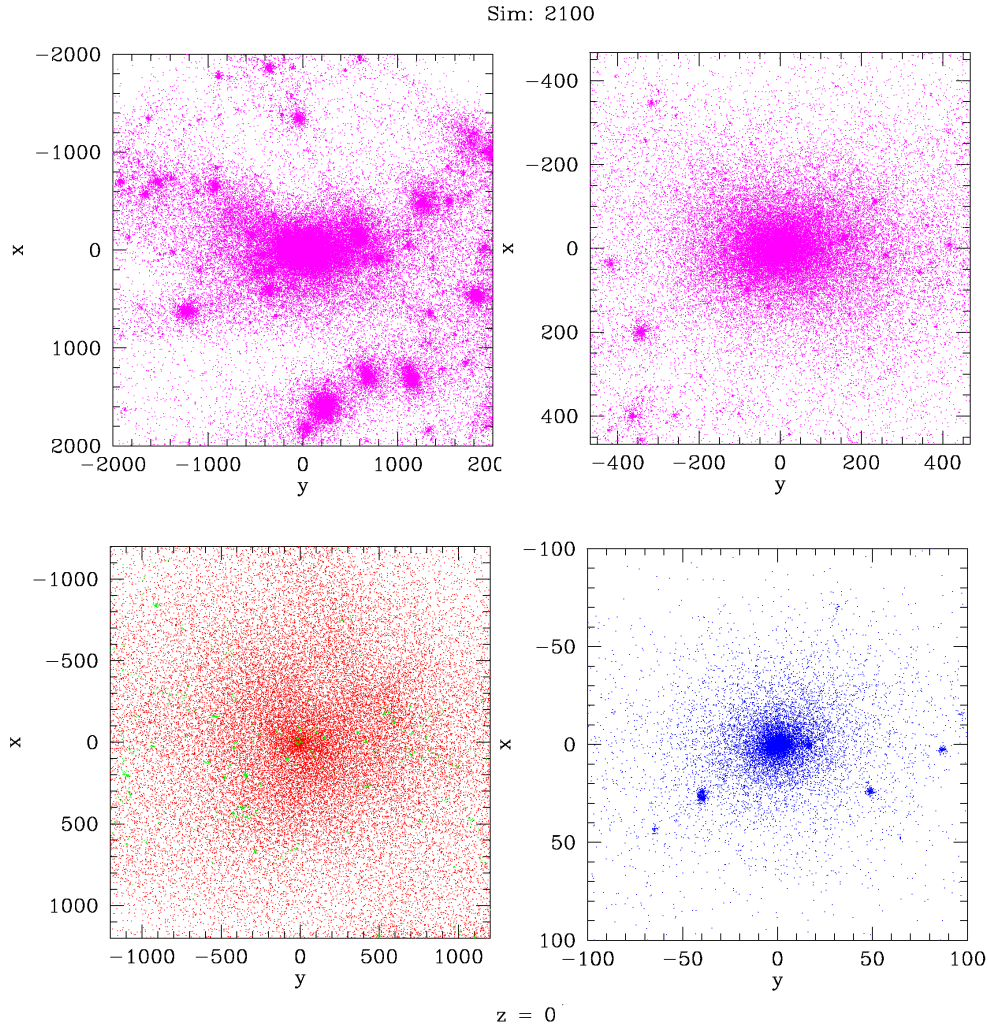}
\caption[]{Projections onto the same plane of the positions of
particles within cubes
centred at the centre of mass of MGO \#1 at $z=0$. From left to right and from
top to  bottom we show the large-scale dark matter distribution,
the  dark matter distribution at the scale of the virial radius,
the gaseous component (hot gas corona and cold clouds) and the stellar component.
Colour codes: dark matter is magenta; hot gas ($T > 8 \times 10^4$) is red;
cold gas is green; stellar  particles are blue.
The window sizes are  in kpc.}
\label{4windows}
\end{figure}

In Figure~\ref{4windows} we show the projections onto a plane of the positions,
at $z=0$, of the particles within a cube centred at the centre of mass of  
MGO \#1.
The cube  size in kpc is specified in each of the four windows.
From left to right and from
top to  bottom we show the large-scale dark matter distribution,
the  dark matter distribution at the scale of the virial radius,
the hot gas component ($T > 8 \times 10^4$) and the stellar component.
Note in Figure \ref{4windows} that the hot gas component is much less
concentrated than the dark matter component, while the stellar component
sets at the very central regions. Note also that cold gas
clumps (in green) can be seen spatially mixed with the hot gas.

To illustrate mass assembly and star formation
  histories  from
high to low redshift, we visualise the PO($z$)s of the two MGOs
at four  $z$s.
To draw the corresponding  plots
(Figures~\ref{PatternsDark}-\ref{PatternsHotGas}),
we have proceeded  with different steps:
i) Consider MGOs \#1 and \#2
 and their constituent particles of different kinds at $z=0$
(Figure  \ref{4windows}).
Choose a particle type at this $z$: dark matter, stellar, cold
or hot gas particles 
and keep the identity of those whose distance to the MGO centre
of mass satisfies $r < r_{\rm lim, i}$.
We used $r_{\rm lim, i} = 160$ and $190$ kpc for $i=$dark matter
in MGOs \#1 and \#2, respectively\footnote{these are the radii enclosing
half their total mass, including dark matter},
$r_{\rm lim, i} = 30$ for $i=$stellar mass component, and
$r_{\rm lim, i} = 1200$ and $1500$  kpc, for $i=$ hot gas component.
ii) Look for the positions of these particles at four different
redshifts, $z' = 6.0, 3.5, 2.2 $ and $1.0$.
 Look for the type (either gaseous
or stellar) of those particles that at $z = 0$ are stellar particles,
as well as the temperature $T(z')$  of those particles that at given $z'$
are gaseous particles.
iii) Finally, to draw the Figures in this  set, project these positions onto a given plane,
the same at any $z'$, and around the same centre at each $z'$,
using the same colour code
as that in Figure \ref{4windows}.

 \subsection{ Hydrodynamic Simulations at  Large Scales:  an
 Outline}
\label{EvoBoxScale}

  As other authors',
  our simulations show that gas roughly follows  dark matter at
  scales larger than a few Mpcs.
As predicted by the adhesion  model  and  expected from
pure gravitational simulations,
our simulations indicate  that at high redshifts
the cosmic web dynamics proceeds as usual. Indeed,
evolution causes first the formation of
small pancakes that, later on, grow and  join, forming a
   three dimensional  cellular network,
   where dense walls and filaments
   surround underdense regions or voids.

Using three dimensional  visualisation
techniques, we have  seen that matter (dark or baryonic) that first sticks into  pancakes moves later on into the filaments,
forming  a nearly irrotational flow, until it comes to the nodes.
Moreover, secondary filaments form into walls, and secondary
nodes form within filaments.
At pancake, filament and node  locations, the velocity field becomes
discontinuous and the density field shows very pronounced maxima
(caustic locations).
These behaviours  are illustrated in Figures \ref{PatternsDark}
to \ref{PatternsBar2}.
Nodes
play the role of local attraction basins where mass piles
up
(this is  shocked material   in the terminology
of the adhesion model).
When caustics form,  volume contractions are
extremely fast and mass becomes trapped,
 resulting in energy dissipation.
 At the same time,
 the gaseous component trapped into singularities begins to be gradually transformed
 into stars at the densest locations.

   This cellular structure
   is not homogeneous, as the mesh size
   and the coalescence length
   $L_{\rm c}(t)$
   depend on position at a given time. And so, some
   subvolumes enclose denser parts of the  cosmic
   network than others, and are
   overdense relative to the average box  density while other subvolumes
   are underdense. Their evolution is quite different ({\it locality} of evolution).
	 In the simulations we see that
   underdense regions expand for ever.
   Overdense subvolumes, at a given scale, first expand slower than average.
   Then their expansion stops, i.e., they turn around, in the
   language of the spherical collapse scenario. Finally, these
   subvolumes experience fast global contractive deformations,
   as the adhesion model  indicates (see $\S$\ref{Hypo}),
   within collapse time-scales. These fast contractions  involve the cellular structure
elements the subvolumes enclose, and, in particular, nodes.
  These nodes are connected by filaments and experience fast  head-on fusions
as a consequence of subvolume contractions.

We have seen in our simulations that  overdense subvolumes
at different scales act as {\it flow convergence regions} (FCRs),
 i.e. attraction basins for mass flows
 with defined boundaries in Lagrangian coordinates, whose
fate is to shrink following contractive deformations
(recall the short-scale smoothing character
of the advanced stages of non-linear gravitational evolution).
The POR of  a given massive object encloses several FCRs, 
which gradually disappear giving clumps at FCRs of a higher
hierarchical level, as visualised and explained in $\S$\ref{ELO1MassAss} below.
Each FCR hosts several nodes along with their
connecting  filaments, where star formation is already on.
At the same time, as predicted  by Burgers' equation,
a certain fraction of gas is never trapped into caustics
and it forms a diffuse phase component that fills FCRs
unconnected to cold gas. This diffuse gas phase is never  involved in
star formation and it gets gravitationally heated at violent events.

From a global point of view, we have also seen that
from the very beginning different FCRs,
corresponding to the PORs of different massive objects,
appear in the whole simulation box, and that some of them merge along the evolution.
An important point is that FCR properties result from the structure of  the minima
of the potential function, $-\Phi_0(\textit{\bf q})$, that is, they are determined from the very
beginning.   
A consequence is that the amount of mass (either dark or gaseous)
available to form an object is fixed ab initio.

\subsection{ Caustic Tree Development vs Regular Particles}
\label{ELO1MassAss}

\subsubsection{Dark Matter and Star PORs at Different $z$s}

\begin{figure}
\begin{center}
\includegraphics[width=.8\textwidth]{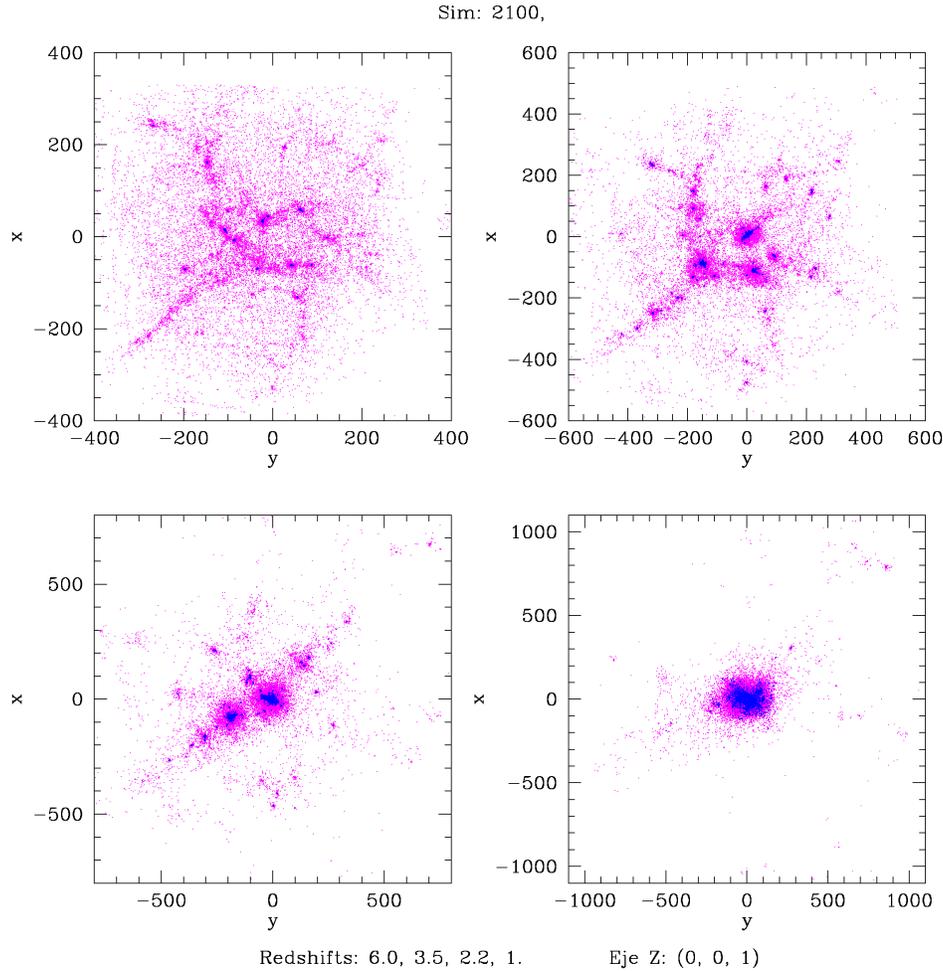}
\caption[]{
This Figure 
visualises the $ PO(z)$ of MGO \#1 at different $z$s.
It shows the projections of the dark matter
particles that at z = 0 form half the more bounded particles
of the halo.
From left to right and from
top to bottom, the redshifts are $z=6, 3.5, 2.2$ and $1.0$.
Colour code is as in Figure 1. The pieces of the cosmic web
displaying different FCRs   clearly stand out at high $z$.
}
\label{PatternsDark}
\end{center}
\end{figure}

\begin{figure}
\begin{center}
\includegraphics[width=.8\textwidth]{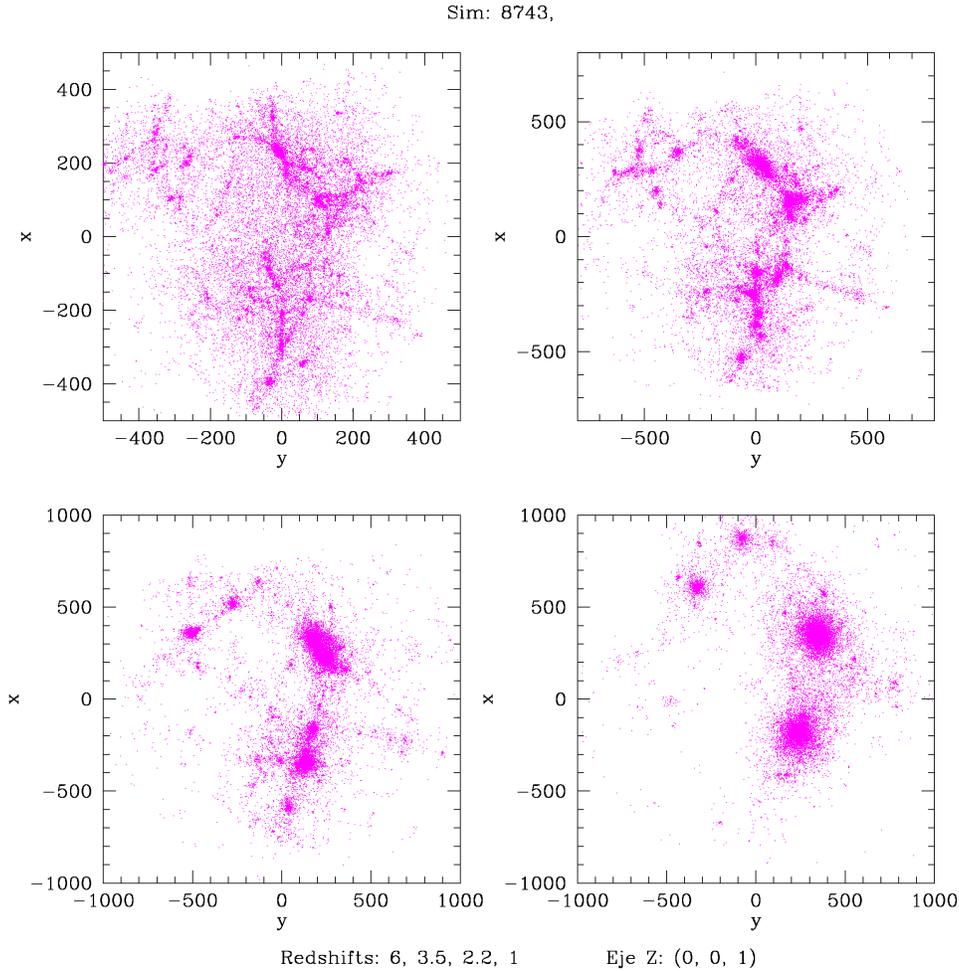}
\caption[]{Same as the previous Figure for the dark matter particles of MGO \#2.
From left to right and from
top to bottom, the redshifts are $z=6, 3.5, 2.2$ and $1.0$.
Different FCRs are clearly apparent as well as their
transformations. Between $z = 1$ and $z = 0$, their
resulting clumps will merge giving the final object. 
}
\label{PatternsDark2}
\end{center}
\end{figure}

\begin{figure}
\begin{center}
\includegraphics[width=.8\textwidth]{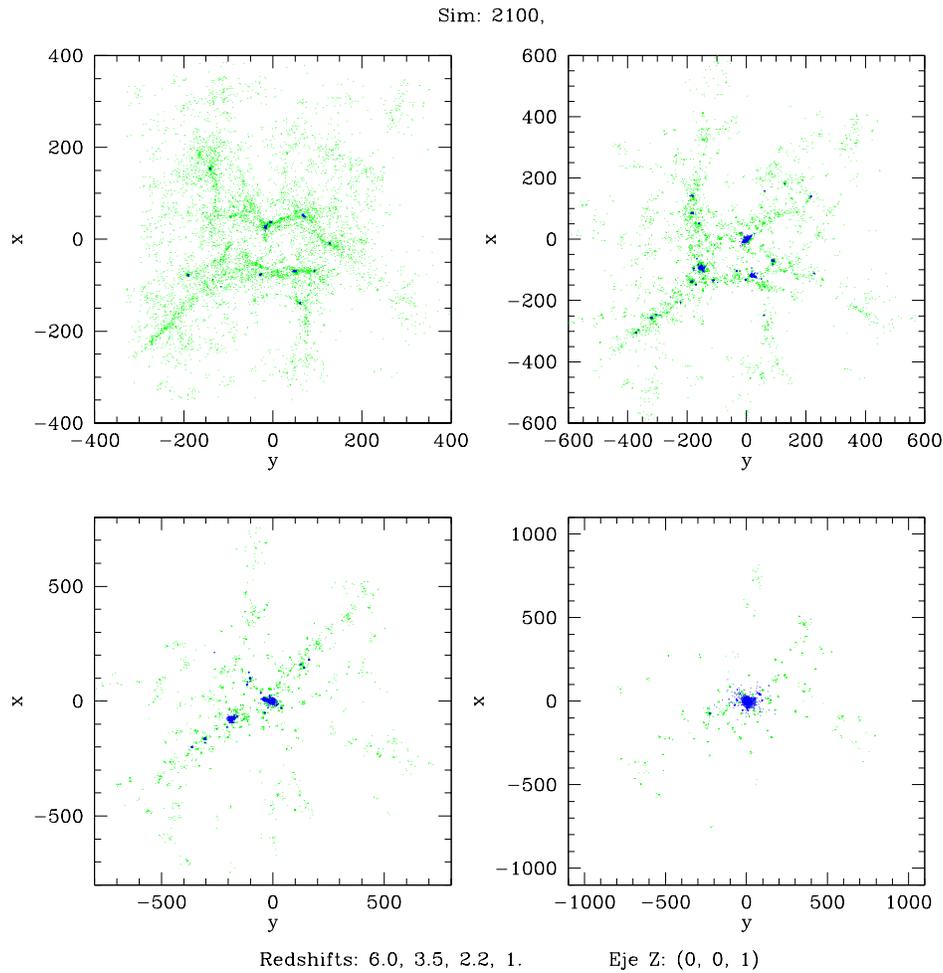}
\caption[]{ The $PO(z)$ corresponding to the stellar component of MGO \#1 at four different $z$s.
This Figure shows the projections of the baryonic
particles that at z = 0 form the stars of  MGO \#1. 
From left to right and from
top to bottom, the redshifts are $z=6, 3.5, 2.2$ and $1.0$.
Colour code is as in previous Figures.
}
\label{PatternsBar}
\end{center}
\end{figure}

\begin{figure}
\begin{center}
\includegraphics[width=.8\textwidth]{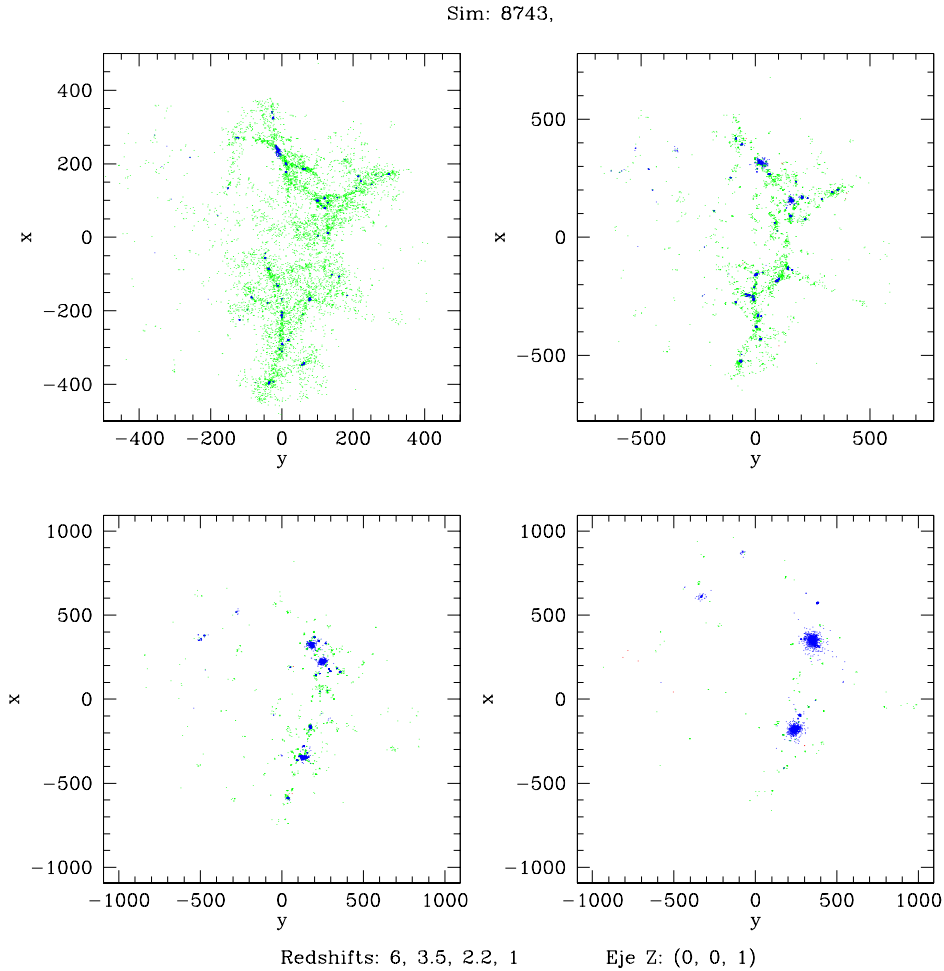}
\caption[]{ Same as previous Figure for MGO \#2.
From left to right and from
top to bottom, the redshifts are $z=6, 3.5, 2.2$ and $1.0$.
Again, different FCRs and their transformations
 clearly stand out, as well as the  galactic objects they produce
by $z=1$.}
\label{PatternsBar2}
\end{center}
\end{figure}

As an illustration of the caustic tree concept, as well as of other ideas
introduced in $\S$\ref{Hypo},
we visualise in Figures \ref{PatternsDark} and \ref{PatternsDark2}
the $PO(z)$s at four different $z$s corresponding to the dark matter particles identified
at $z_{low}=0$.
In Figures \ref{PatternsBar} and \ref{PatternsBar2}
we visualise the $PO(z)$s corresponding to 
the baryonic matter particles that
at $z_{low}=0$ end up as
the stellar component of these two MGOs.
As stated above, caustics (the points where singularities
build up) stand out in these Figures at $z=6$ and $z=3.5$ as the regions where
mass piles up, and in fact they are the densest parts of
the cosmic web that shows up  at these scales. A diffuse component is also evident,
and constitutes the regular points at the corresponding $z$s. Note that
this diffuse component tends to flow into caustics, therefore vanishing as evolution proceeds.

We can see in these Figures that at high $z$
the evolution patterns of the dark matter and baryonic
components closely follow each other.
In fact, at $z=6$, 
singularities build up
roughly at the same locations in either of these two components.
Note also that stars (in blue) are formed
from gas located in small clumps at the denser subvolumes of the system, as sampled
either by dark matter or baryons. From $z=6$ to $z=3.5$,
and for either MGO, contractive 
deformations, acting on different FCRs, bring most baryons into a 
more marked outstanding filamentary structure. During this process
star formation goes on at the densest subvolumes of this filamentary structure.
Some of these clumps merge 
during this period.
The dark matter component is also brought by the contractive deformations 
into more marked filaments. However, a 
difference with the baryonic component stands out at $z=3.5$:
caustics are thicker in the case of dark matter, which is non-dissipative.
Baryons in turn, have dissipative behaviour, and become denser in 
phase space as they cool.

We now consider the evolutionary patterns of MGO \#1 and MGO \#2
between $z=3.5$ and $z=2.2$.
We see that at $z=2.2$ filaments have practically been removed 
in favour of clumps, either in the dark matter or the baryonic components
(again as a result of contractive deformations).
In either MGO \#1 or MGO \#2,  a number of massive clumps stand out
at $z=2.2$.
The dark matter halos of these clumps are much more
massive and bigger than the baryonic component they host, which is concentrated
due to dissipation, and mostly stellar.
Note that some of these dark matter halos are about to merge, while
their baryonic components are not yet involved in a merging process.
In the $z=2.2$ snapshot, several FCRs can still be seen
around the massive clumps, but are now dominated by clumps.

Contraction goes on from $z=2.2$ to $z=1.0$.
At $z=1.0$, filaments in both the dark matter and the baryonic components
have completely disappeared and  dark matter relaxed spheroids
have formed  (almost relaxed in the  case of MGO \#1). Note that
one such spheroid forms at each FCR seen in the previous snapshot.
These dark matter spheroids host  most of the baryons  that at $z = 0$ 
form the stellar component of either  MGO \#1 and MGO \#2.
Note also that most of these baryons have already been transformed into stars.
The stellar component of MGO \#1 is involved in a major merger event
at $z=1.0$, resulting in the object depicted in Figure \ref{4windows}.
Regarding MGO \#2, the four spheroids seen at $z=1.0$  will eventually merge
between $z = 1$  and $z = 0$, resulting in the final object at $z=0$.

 A remarkable fact is that no hot gas particles
(in red) show up in these Figures.
Otherwise, it is also very remarkable that the assembly patterns
for either MGO \#1 or \#2  are quite similar,
both for their dark and stellar components.
We remind that S2100 and S8743 have been run with different
codes (indeed, the DEVA code has been extensively modified during its OpenMP parallelisation),
different box sizes and different values of the cosmological
model or the SF parameters.
These patterns are common to those we have found for the other
massive galaxies  in S2100 and S8743, or any other simulation we have
analysed.

Note that caustic formation
implies different evolution patterns for gas and dark matter at
small scales. Otherwise, cooling processes dominate at  the densest filaments and
nodes, where the gas is cold, while 
hot gas flows at FCRs end up as 
a hot corona around the main clumps.

\subsubsection{Hot Gas}

\begin{figure}
\begin{center}
\includegraphics[width=.8\textwidth]{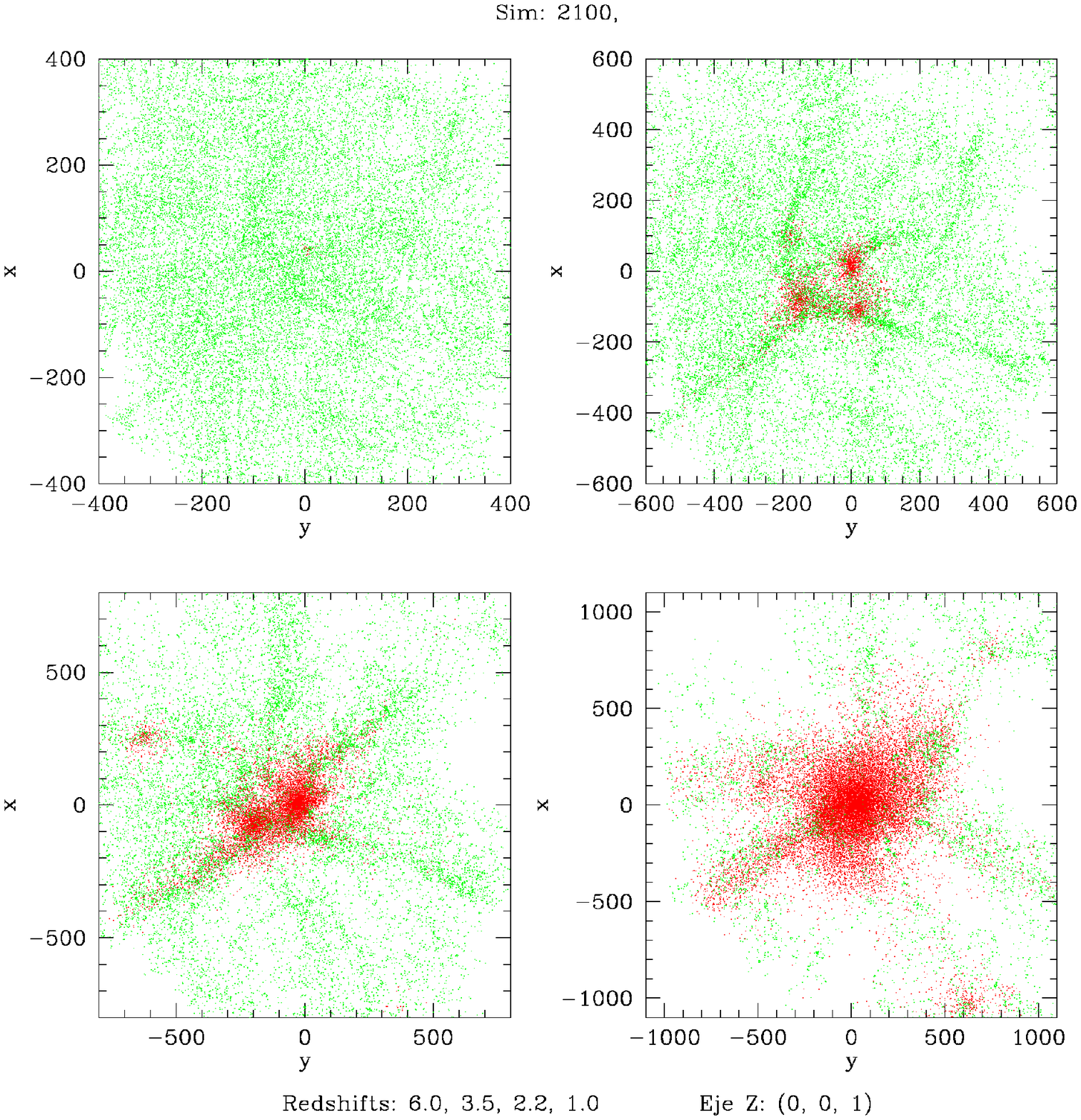}
\caption[]{The $PO(z)$ of a massive MGO at four different $z$s.
This Figure shows the projections of the 
gaseous particles that at z = 0 form the hot corona of  MGO \#1.
From left to right and from
top to bottom, the redshifts are $z=6, 3.5, 2.2$ and $1.0$.
Colour code is as in previous Figures (red:  gaseous particles with $T > 8 \times 10^4$K;
green: gaseous particles with $T \le 8 \times 10^4$K).
}
\label{PatternsHotGas}
\end{center}
\end{figure}

We now turn to analyse the origin of the hot gas component shown in
Figure \ref{4windows}.
We focus on the MGO \#1 hot gas component, because the patterns
for  MGO \#2 or the other massive galaxies formed in any simulation
we have analysed are quite similar.
In Figure \ref{PatternsHotGas} we show four snapshots, at the same
$z$s than the  previous Figures, of those gaseous particles
that at $z = 0$ are within $r < r_{\rm lim } = 1200$ kpc
of MGO \#1 centre and
have temperatures $T > 8\times 10^4$K. In this  Figure,
green points are cold gas  ($T \le 8 \times 10^4$K)
while red points are hot gas at the $z$ considered.
We can clearly witness in this Figure the gravitational gas heating
due to violent dynamical events, that partially transform the mechanical
energy involved into contractions into thermal energy.
 To be quantitative,
recall for example that a system must get rid of an amount of energy equal to its 
binding energy as it collapses from infinity  and
 virialises, see \citet{Binney:2008}.
Hence, gas heating first begins at the densest subvolumes within
the system. Another outstanding characteristic is that contractive
deformations have quite a different effect on this component than
in the dark matter or stellar components. Indeed, even if diffuse gas is pulled
by dark matter and tends to move towards the same zones where dark matter does,
the total volume they fill at $z = 6$ has not had its shape significantly
deformed between $z = 6$ and $z = 2.2$. In other words, 
$PO(z)_{hot, gas}$ suffers from less shrinkage and shape changes than 
$PO(z)_{cold, gas}$ or $PO(z)_{dark, matter}$
(compare Figures \ref{PatternsDark} and \ref{PatternsBar} with Figure \ref{PatternsHotGas}
at $z=$ 6 and 3.5).  
Note that this gas has a much lower phase space density at any $z$ than
that of dark matter particles or baryons that eventually end up
as the stellar MGO component.
Between $z = 3.5 $ and $z = 2.2$, diffuse gas follows the collapse of
 the denser
subvolumes of the configuration and heats.
From  $z = 2.2$ to  $z = 1.0$,  
while the two halos seen in the $z = 2.2$ snapshot complete their merging,
gas heats and tends to diffuse due to 
gas pressure and heating, and mechanical energy conservation.
Finally, between $ z = 1.0$ and $z=0$,
heating and diffusion keep on acting, until 
gas  forms the hot corona around   
MGO \#1 seen in Figure \ref{4windows}.
This behaviour of the hot gas component illustrates the concept
of regular mass elements at a given redshift, defined
as those that have not yet been trapped  into
a singularity at this redshift.
They are less dense than those that have been trapped.
This is the case for the gas in this  Figure as compared to the
gas providing the mass to form stars (see previous Figures).
Remember that the existence of singular and regular mass
elements  is predicted by  Burgers' equation solutions
(see $\S$\ref{Hypo}).
In fact, this mathematical distinction between singular and regular mass elements is at
the basis of the different physical behaviour we have found 
between 
dense gas and diffuse gas in the simulations.
Note that a fraction of the gas that transforms into stars is 
heated  as it gets into the object or comes close to it.
However, it cools very fast due to its high density \citep{TesisJose}, see also
O\~norbe et al. 2011.

\subsection{Evolution at the MGO Scale: Star Formation}
\label{EvoELOScale}

We now consider the formation and ageing of
stars in the same objects as in $\S$\ref{ELO1MassAss}.
The aim is to  assess details of stellar  formation as a consequence
of dynamical activity and their configuration changes
(i.e., irregular shapes, relaxation).
To this end, we  focus on processes at the scales that are now important
($\sim$ lower than 50 kpc).

\begin{figure}
\begin{center}
\includegraphics[width=.8\textwidth]{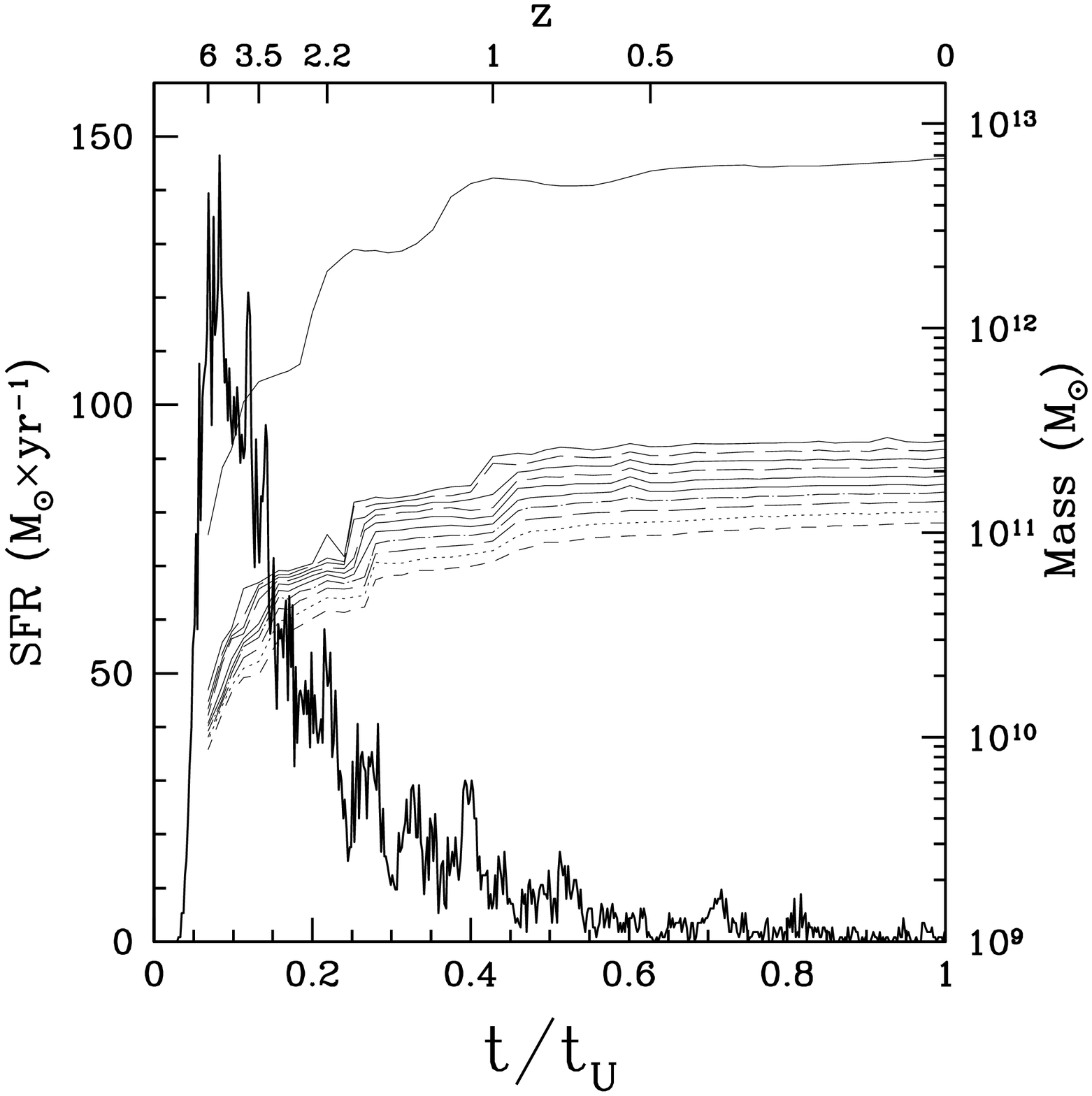}
\caption[]{The star formation rate history (solid line) for the same stars drawn in Figure 
\ref{4windows} and the mass aggregation
track  along the main branch of the
merger tree for MGO \#1.
Upper thin full line: the virial mass.
Point and dashed lines below: baryonic mass inside different fixed radii, equally spaced in a logarithmic basis
 from 3 to 30 kpc. 
}
\label{MAT1}
\end{center}
\end{figure}

\begin{figure}
\begin{center}
\includegraphics[width=.8\textwidth]{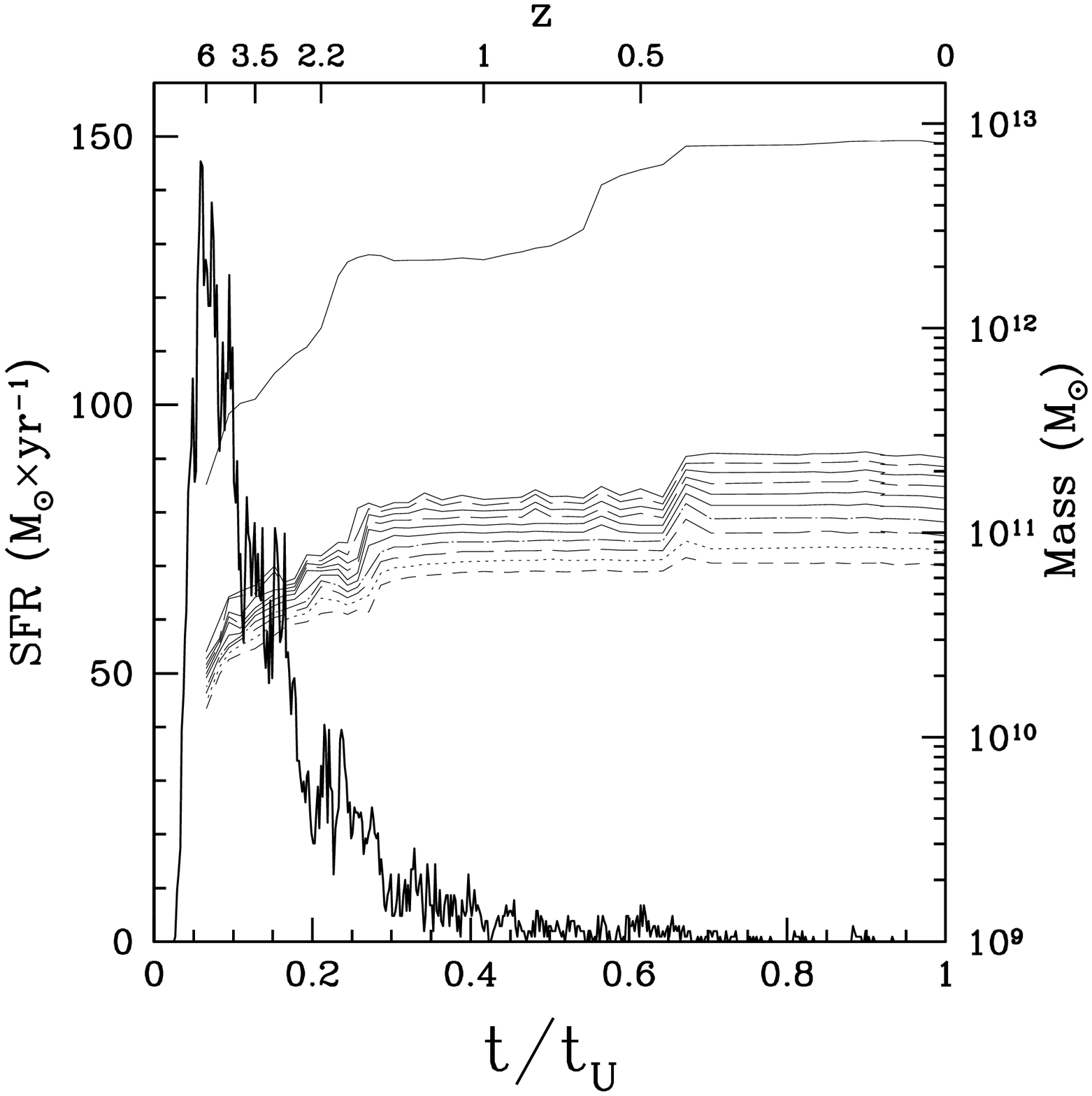}
\caption[]{Same as previous Figure for MGO \#2.
}
\label{MAT2}
\end{center}
\end{figure}

We have already seen that at high  redshifts, stars first form
in the densest and coldest gaseous nodes. As
evolution proceeds, 
flow convergence regions become denser and denser, until they
experience global contractive deformations.
As seen in Figures \ref{PatternsBar} and \ref{PatternsBar2},
such  contractive
deformations involve the  FCRs  of either MGO \#1 or \#2
from z$ \approx 6\, (t/t_U = 0.068$) to z$ \approx 2.2\, (t/t_U = 0.22$),
 triggering enormous activity and causing a very fast
coalescence of the different FCRs. Indeed, the nodes
they host become
closer and closer until they hierarchically merge with very low
relative angular momentum. This results in very modest orbital
delay, which in turn leads to high rates of gas energy dissipation and
important bursts of star formation at the nodes. 
The SFR histories of the stellar population present at $z=0$ 
 have been plotted in Figures \ref{MAT1} and \ref{MAT2}. 
They indicate that most stars ($\approx $70 \%)
 were formed when the Universe 
was younger  than 20\% of its present age.

By $z \approx $2
the dynamical activity slows down; the inflow rates decrease and
most previously flowing mass is now stuck onto dense small objects
(i.e., secondary nodes) that later on
fall towards the central  object after orbiting it as satellites.
Any further MGO  mass assembly takes place
through merger events with 
non-negligible relative angular momentum
and longer time-scales, involving either small satellites or almost
equal-mass objects.

To illustrate this point, the mass aggregation tracks 
along the main branch of the merger tree have been drawn for
the MGOs,  for both the baryonic component
(mass inside fixed radii) and for its total mass (virial mass)
in Figures \ref{MAT1} and \ref{MAT2}.
These mass aggregation tracks give us information on the mass assembly processes through time.
Major mergers ($M_{secondary}/M_{main} > 0.25$), minor mergers  
and slow accretion processes in the dark matter or baryonic component
can be clearly identified as discontinuities
or slow mass increments, respectively.

The dynamical slowing down of MGO mass assembly 
 at $t/t_U \approx 0.2$ can be clearly 
appreciated in the mass aggregation tracks (dark matter component),
that  also causes the MGO  baryonic component assembly to slow down. 
 In some cases, 
the time delay between halo and  baryonic component coalescence
at these lower $z$s can also be appreciated.  This corresponds to an orbiting infall
of the merging object around the main.

  Stellar objects have
  irregular shapes while  violent dynamical  events are ongoing.
Afterwards, as the rate of mass assembly slows
  down, the main object progressively acquires a
  regular shape, as corresponds to higher and higher degrees of
  relaxation. A passive ageing of its stellar population can
also be appreciated. Gas accretion in this period occurs
  at a slow rate, while star formation at the centre of the
  proto-MGO (at a very low but continuous rate)
  leads to a space segregation of stellar ages.

 The first phase of this process,
 corresponding to the contractive deformation, 
 is termed {\it fast multiclump collapse}.
 Some authors  have heuristically
 proposed \citep{Thomas:1999} this concept to account for the need
 of short SF time-scales at high $z$  in early type  galaxies within
 the hierarchical clustering paradigm.
 The new point here is that fast multiclump collapse
 {\it directly results from the simple physics}
 involved in our simulations of the evolution
 of perturbations to a $\Lambda$CDM model.
In fact, it is a natural and important consequence
of the  adhesion model we use as a guideline to  highlight this underlying physics.
 The second phase at lower redshift is dominated by 
 classical mergers with a lower rate of mass aggregation.

\subsection{Two Phase Mass Assembly Process for Massive Galaxies}
\label{twophases}

  The mass assembly histories of the particular MGOs outlined in the
previous sections are representative of the main patterns present in the general
MGO population.

  Analytical models, as well as N-body simulations,
show that two different phases can be distinguished along
{\it halo}  mass assembly \citep{Salvador:2005,Wechsler:2002, 
Zhao:2003}:
i) first, a violent,  fast phase, with  high mass aggregation 
(i.e., merger) rates; and
ii) later on,  a slow phase, where  the  mass aggregation rates
are much lower. Previous smaller box hydrodynamical simulations had already  confirmed this scenario
and its implications on MGO properties at low $z$, see
\citet{DTal:2006}, see also  \citet{Oser:2010} and \citet{Cook:2009}. 
The present analysis, including larger box hydrodynamical
simulations, provide information about the consequences of such
a scenario on the baryonic component
during MGO assembly.

  As noted above, during the fast phase at high $z$, mergers 
are induced by the collapse of FCRs. They are usually multiple,
and involve only low or very low relative orbital 
momentum\footnote{The merger characteristics
have been detected and measured using three dimensional visualisation techniques, as well as
the mass aggregation trees of the objects \citep[for details see][]{GonzalezGarcia:2009}.}.
They are gas rich, so that most of the thermal dissipation occurring
along the mass assembly of a given MGO takes place during this phase,
and results in strong SF bursts that transform most of the gas available
at the FCR at high $z$ into stars.

 Later on, during the slow phase, mergers result from 
 coalescence of basins, including the main object and its satellites,
 as well as the dark matter halo and the diffuse hot gas corona.
Mergers now happen
with relative orbital  momentum so that there is a time delay
between the halo fusion and the MGO fusion  due to orbital decay.
This can be appreciated in Figures \ref{MAT1} and \ref{MAT2}
when the aggregation tracks for dark matter and baryons around
major mergers are compared, or in the projected images of MGO \#1 and \#2 at $z=2.2$.
Their rate is much slower, and it can be even  zero; that is,
some MGO have not experienced any major merger event during
this second phase. The actual rate depends on the density of the 
environment where the MGO lives. 
Because most gas available at the POR is exhausted during the previous fast phase,
the dissipation and SF rates are usually low in this phase, even when
a major merger occurs (see Figure \ref{MAT1} at $t/t_U \approx 0.4$,
or \ref{MAT2} at $t/t_U \approx 0.65$
where  major mergers take place  with only very modest consequences on the SFR).

 These two phases of mass assembly are responsible for different
properties of MGO samples, see for example \citet{DTal:2008}. This will be further developed in a forthcoming paper.

\section{Summary, Discussion and Conclusions}
\label{implications}

\subsection{Summary on Simulation Results Confronted with Singular Flow Properties}
\label{Confronted}

The results presented in the previous section help us decipher 
massive galaxy formation in the light of some generic properties
of singular flows, as described by the solutions  of Burgers' equation.  
Our most significant  result is that massive galaxies form
within {\it  flow convergence regions} (FCRs)
as a result of the non-linear evolution of the
primordial density fluctuation field.
As seen in the simulations, FCRs are overdense regions that act as attraction basins for mass flows
and undergo contractive deformations transforming them into nodes of larger scale FCRs
(i.e., at a superior level of the hierarchy).

Interpreted within the theoretical framework put forward in $\S$\ref{Hypo},
our simulations suggest that to form a massive  galaxy, three conditions must be met:
(i) its caustic tree  must include
larger shocks than average; (ii) the mass contained in its POR, $PO(z_{in})$,
must be of the order of the virial mass of a massive elliptical galaxy; and (iii)
its POR must contain very deep minima of $-\Phi_0(\textit{\bf q})$\footnote{Note that this can be considered
as a reformulation of what \citet{Evrard:1990} propound, and
that PORs fulfilling these requirements are more frequent
in  subregions of the simulation that are overdense from the beginning.}. 
We remind that FCR properties, as those of solutions  of Burgers' equation,
 result from the structure of  the minima
of the potential function, $-\Phi_0(\textit{\bf q})$, that is, they are determined from the very
beginning. These properties include where massive FCRs of different scales are  localised within
the simulation box, the Universe age when they collapse, the time-scale for their collapse
and the mass they enclose, among others.

According  to the time and space scale invariance properties of  Burgers' equation, in the absence of energy exchanges (heating and cooling), all caustic  trees are similar in a statistical sense,   
once conveniently rescaled in space and time variables, see $\S$\ref{Hypo} and \citet{Vergassola:1994}.
 In this case, only one parameter would determine the difference: the mass each 
caustic tree involves, that is, the mass enclosed by the POR of the object,
equal to the sum of the masses of the constituent FCRs. 
An important point is that mass assembly and star
formation can  be slowed down by angular momentum.
Hence, FCRs where massive galaxies get their mass assembled must be
relatively free of vorticity. This  condition might not be difficult
to be met  ab initio in massive
FCRs in the adhesion model, where the velocity field is potential.
Moreover, as massive FCRs collapse at high $z$, there might not be enough time for vorticity
to build up before   their collapse  \citep{Buchert:2005,Knebe:2006,Governato:2004}.

As a consequence of the conditions above,  in the regions $R$ defined by the $PO(z)$ of massive galaxies,
the coalescence length $L_c(R, t)$  grows faster than average.
Remember that $L_c(R, t)$ is defined in such a way that 
coalescence dominates and substructure is substantially erased 
at scales  smaller than it, at time $t$ and within region $R$ (see $\S$\ref{Hypo}). 
Hence, within these regions, the caustic tree is travelled through very fast
and  the collapse at  given  scales  occurs earlier on than average
at these scales.
Indeed, our simulations show that a lot of caustic formation  takes place at $R$ very soon,
including    mergers of nodes  at very low angular momentum,
that is, without any significant orbital delay (short time-scales).
The mass inside $R$ gets  virialised
when the  caustic formation event or collapse associated with  the
deepest minima of $-\Phi_0(\textit{\bf q})$ at $PO(z_{in})$ takes
place at time $t_f$.
Previously, there had been  lots of activity involving caustics
 of different geometries and on  smaller space scales. This activity is
 associated with less deep minima of $-\Phi_0(\textit{\bf q})$,
 and  with their evolution through the swallowing up of  cells. 
At $t_f$ the violent collapse event involves the  structures resulting from
previous  caustic formation events
at  smaller scales within $R$.

Ideally, each one of these events takes the mass (including gas) 
of the volume they involve  into
a lower dimensional subvolume (remember the definition of
flow singularities in terms of the
Lagrangian map given in $\S$\ref{Hypo}). However,
simulations indicate that the singularities get indeed dressed.
Simulations results show that most  particles involved in caustic formation live
in much smaller,  shrunk, but  three-dimensional regions.
In any case, 
the violent event at $t_f$ associated with
the deepest minima of $-\Phi_0(\textit{\bf q})$ at $PO(z_{in})$ tends to
 clean the mass around it, and more so when it involves a large volume.
Hence, the amount of mass available to be further accreted after $t_f$ is severely
limited. This would explain why the mass accretion rate is in many cases drastically reduced 
afterwards.

Apart from mass involved into singularities,
 Burgers' equation solutions ensure the existence
 at any time $t$ of regular points or mass elements,
 as those that have not yet been trapped into a singularity
 at $t$ ($\S$\ref{Hypo}).
 This is confirmed by the simulations, and allows us to understand
 the presence of a large-scale diffuse, hot gas component.
 This component has been
 gravitationally and hydrodynamically heated in violent events
 (either collapse or contractive deformations at high $z$, and major mergers
 at  lower $z$ values) and it carries a fraction of the mechanical macroscopic
 energy and momentum involved in these events. Remember that to assemble
a mass  $M_{\rm vir}$, the system has to get rid of an amount of energy equal to
its binding energy, which per unit mass is proportional to $M_{\rm vir}^{2/3}$. 
So, even when thermal energy exchanges are allowed for,
$M_{\rm vir}$ is a key parameter.

We have presented a global point of view for the process of mass assembly. 
From the point of view of the massive object that is being assembled, our simulations indicate that,
at high $z$, MGOs
get their baryons  either through almost equal-mass mergers
of subunits made out of gas and stars and connected by filaments,
or through smoother gas accretion along filaments.
Most accreted gas is cold and dense, and it has been previously
involved in caustic (pancake and filament) formation,
while at any $z$ a fraction of gas or dark matter has never been involved in this kind
of events.

\subsection{Some Generic Observable Implications}

Some currently available observations
of massive ellipticals can be explained in this
 scenario of {\it  flow convergence regions}.
 Here we  focus on the most generic and physically relevant consequences
 of this scenario, at a rather qualitative level.
We postpone the quantitative  discussion of more detailed,
 observationally related consequences to a future paper.

 \begin{enumerate}
 \item
The existence of regular mass elements  (i.e., a diffuse gas component)
is assured  by Burger's equation. Due to its low density,
 this diffuse component cannot cool rapidly, and remains as a
 high-frequency radiation emitter (i.e., X-rays in massive
 ellipticals or groups). Moreover, most hot particles do not fall towards the centre of the MGO
configuration, but their distribution is at the scale of the
virial radii or beyond \citep{Onorbe:2007}.

\item
From a global point of view, the evolution of high-$z$ mass density elements
(either dark or baryonic)
consists in  their  accumulation within attraction basins
around flow singularities (except for regular mass elements). Due to dissipative
processes, the tendency to accumulate is even stronger in the case
of baryons belonging to caustic trees. In this context, black hole formation at high $z$
at the points of flow convergence (the centres of elliptical
galaxies or of spiral bulges) appears quite natural.
Their later growth in a scenario of co-evolution for spheroid and
black hole mass accretion, fed by gas inflows
from their FCRs, and self-regulated at smaller scales by more complex
astrophysical local processes \citep[see for example,][and references therein]{Granato:2004,Silk:2010}  
could explain many of the correlations
black holes and their hosts show. A  tantalising interpretation is that
black holes at the centres of massive galaxies come  from  {\it node-like flow singularities},
while the massive galaxies correspond to their {\it dressing}.

\item
We now address the role of  $M_{\rm vir}$  to build up correlations.
$M_{\rm vir}$ is the halo total mass-scale once the object gets bound,
and the mass enclosed at its POR (or $PO(z)$)
before this happens.  To make its role clear, sometimes it is easier
to talk about the mass corresponding to the massive FCRs that end up
in the bound object at low $z$.

We first address correlations involving the stellar mass of the massive galaxies, $M_{\rm bo}^{\rm star}$.
Stars form at shocks ensuing caustic formation. 
According to $\S$\ref{CodeSF}, the  gas mass fraction susceptible
to be transformed into stars in a given caustic event is equal
to the fraction of gas involved in this event
whose  density is  higher than the threshold density $\rho_{\rm  thres}$.
According to the Kennicutt-Schmidt-like law implemented in  our simulations,
those gaseous elements with  densities above the  threshold
are  subsequently  converted into stars with a
time-scale  set by  $t_g/c_{\ast}$,  where $c_{\ast}$  is
the star formation efficiency parameter.
For massive galaxies, the time-scale must be  short so that
most gas elements above the density threshold are rapidly
transformed into stars.
In a FCR, the gaseous mass 
involved in a caustic event is roughly proportional to 
that of dark matter. Hence, the stellar mass produced at each caustic
event is roughly proportional to the dark matter it involves.
Otherwise, the stellar mass of an object at a given redshift $z1$
is the sum of the stars produced in each caustic event of its
caustic tree at $z1$. 
These facts explain the correlation found between the intrinsic
parameters $M_{\rm vir}$ and $M_{\rm bo}^{\rm star}$,
and indicate that the correlations are primarily with
$M_{\rm vir}$ for the intrinsic objects, 
see \citet{Onorbe:2005} and \citet{TesisJose}.
Moreover,
as explained above, the amount of energy a system has to get rid of 
per unit mass to bound a mass $M_{\rm vir}$ is proportional to $M_{\rm vir}^{2/3}$.
This is in part inverted into dynamical heating, that is,
velocity dispersion. This explains the correlation
found between $M_{\rm vir}$ and the velocity dispersion at
different scales, see \citet{Onorbe:2005}. 

Now, as $M_{\rm bo}^{\rm star}$ is proportional to
luminosity L for ellipticals \citep{Kauffmann:2003a,Kauffmann:2003b},
we infer that $M_{\rm vir}$ should be correlated with
luminosity too. As $M_{\rm vir}$ is also correlated with
$\sigma_{\rm los, 0}$, we deduce that L and $\sigma_{\rm los, 0}$
should be also correlated. This is the Faber-Jackson
relation \citep{Faber:1976,Bernardi:2003a}. 
Note, however, that there is a third parameter:
the size. It depends on $M_{\rm vir}$ through
the virial relation, and the tilting of the fundamental
plane due to dissipation, see \citet{Hyde:2009} and references therein. 
For details  on the results of simulations, see
\citet{Onorbe:2005} and \citet{TesisJose}.

\item
Second, we address correlations involving age effects 
of the stellar populations.
Remember that, as we said in $\S$\ref{Intro}, observations suggest that 
the stellar populations of massive galaxies have older mean ages
and lower rates of recent star formation than less massive ones,
and that these effects increase with their mass.
The following considerations 
help to understand them.

As explained before, the amount of gas transformed into
stars at a given time at a given FCR depends on 
how many caustics have formed within it, and the
gaseous mass each of them involves.
Thus, the relative amount of gas transformation depends
on the dynamical activity at the FCR.
As for the dynamical activity, in
those $PO(z)$s defining subregions $R$ with higher mass-scales,
$M_{\rm vir}$,
the local coalescence length $L_c(t, R)$ 
tends to grow faster than
average, and the time unit tends to be locally shorter
than at  other $R'$s where it grows slower. Hence, statistically
the same processes occur at $R$ and at $R'$,
but earlier on, faster  and involving more gaseous mass at $R$.
Consequently,  the tendency is that stars form and the available gas
becomes exhausted earlier on, and on shorter time-scales, within
those PORs whose mass-scale $M_{\rm vir}$ is higher.
This {\it dynamical  downsizing} (i.e., earlier cold  gas consumption
at more massive PORs due to enhanced dynamical activity) is just a consequence  of Burgers' equation.
Of course, this interpretation does not exclude that other
physical or astrophysical processes,  
acting at smaller scales, come into play and
conspire to stop star formation at massive ellipticals,
see discussion for example at \citet{Pipino:2009} and references therein.

\item
Massive old objects in a young Universe  (at redshift $z_{high}$) can be
explained as the consequence of a global contractive deformation acting on
a $PO(z)$ involving much more mass than average, implying that 
it has already collapsed and most of the available gas there has been transformed into
stars at $z_{high}$. Moreover,  if its neighbouring  $PO(z)$s
evolve by themselves, then the MGO would have had time
to dynamically relax  before being disturbed by a violent event.
Both are possible situations in our scenario \citep{Onorbe:2011}.

\item
The  formation scenario  highlighted by our simulations indicates that
 the mass infall rate  onto a given MGO  
is very high   at  the time when its FCR  collapses.
At that time, the gaseous mass elements  either travel up to the very centre, feeding the central
%Tha energy goes in part to gas heating and in part to dynamical heating.
black hole,  or are transformed into stars. This helps to explain that
the SFR history and the AGN luminosity  should be correlated,
i.e., the more active a galaxy is, the younger and the more massive 
the host stellar population tends to be, and,  moreover,
for massive objects both of them peak at high $z$ \citep{Kauff:2003,Falomo:2008}.

\item
Gas in filaments fragments and collapses, forming on their turn
nodes. If they are close to a massive proto-object,
the nodes and the gas  left in the filaments
might be  captured later
 on by the attraction basin of the proto-object.
In this way, the nodes become satellites, that eventually
can merge with the massive object, feeding  starbursts
 depending  on how much dense cold gas
is available. Central starbursts with this origin could explain
blue cores in massive ellipticals, see $\S$\ref{Intro}.

\item
High-$z$ galaxies have in general messy morphologies, with dispersion-dominated
kinematics. 
At lower $z$s, observations suggest that the later the galaxy type and the later
its morphology gets stabilised as $z$ decreases \citep{Zheng:2005,Neichel:2008}.
This can be explained because the  POR mass-scale  determines 
how fast its coalescence length grows on average at PORs of different masses,
that is, how fast gaseous mass availability to be accreted is exhausted.

\item
During the slow phase of mass assembly,  merging is still  taking place,
the stellar mass bound in massive galaxies increases (Gonz\'alez-Garc\'{\i}a et al,
in prep.),
and dry mergers \citep{Conselice:2003}
dominate the mass growth  (see $\S$\ref{Intro} for observational informations).
Thus, the final mass distribution of MGOs, as well as their shapes and kinematical
properties, among others,  is mostly set by 
non-dissipative mergers  \citep{GonzalezGarcia:2003,GonzalezGarcia:2009}.
\end{enumerate}

\subsection{Discussion and Conclusions}
\label{Summary}

In this paper we present a scenario for massive galaxy formation in a
cosmological context 
evinced from hydrodynamical simulations. The theoretical background 
follows from an extension of the {\it adhesion model}
to a situation where dressing of flow singularities is taken into account.
Gas cooling and heating processes, as well as
gas transformation into  stars, are  also allowed for. Moreover,
the locality of flow properties  in an inhomogeneous
density field is also considered.
This scenario shares characteristics of both
the monolithic collapse and  hierarchical  scenarios,
but it is distinct from  them both
and has implications that could  explain generic observational
properties of massive galaxies. 
A two-phase mass assembly clearly shows up in the hydrodynamical 
simulations analysed in this work.

The scenario  summarised in $\S$\ref{Confronted} 
requires that gas is transformed into stars very fast at massive FCRs.
On the contrary, gaseous structures could be formed, including massive disks,
that could survive up to the slow mass assembly phase, and delay massive  
elliptical formation  as a result of mergers of these gaseous systems.
An efficient star formation activity demands that short scale gas accumulation
is not prevented, for example by angular momentum. 
So, gas at these massive FCRs must have a low vorticity.
This is a condition that  might not be difficult to fulfil,
at least is some FCRs, because the motion is potential in the
adhesion approximation. 
Thus, very high gas densities
at the very short scales
where stars are assumed
to form \citep{Henne:2009},
could easily be attained  in an environment  of high dynamical activity.

This scenario for massive galaxy formation revealed by the
simulations disagrees with the conventional picture of galaxy
formation introduced in  the classic papers
by \citet{Rees:1977,White:1978}.
According to these influential papers, all the 
gas in dark matter halos
is shock heated to the halo virial temperature. Then hot gas in
the denser, inner regions of the halo cools as it radiates its
thermal energy \citep[cooling-flow hypothesis,][]{Cowie:1977,Fabian:1977}. The gas
finally settles in a centrifugally supported disk,
where stars form.
Feedback is introduced at this level to avoid that the
infalling gas overcools, and to regulate star formation.
Later on, mergers of stellar disks would produce spheroidal systems.
However, at variance with these expectations, Chandra and XMM
observations indicate much less cooling at the centres of both clusters
and massive ellipticals than expected \citep{Mathews:2003}.
There are also some  indications that gas that
forms galaxies falls in cold and unconnected with
the hot gas \citep{Sparks:1989}.
In consistency with these findings, \citet{Binney:2004}
argues that virial heated gas cannot cool and give
rise to star formation \citep[see also][]{Binney:1977}.
Instead, stars and galaxies would
form out of the significant gas fraction that does not heat
at halo virialisation. This is consistent with our results, 
as well as with those of other SPH simulations
\citep{Kay:2000,Fardal:2001,Katz:2003,Keres:2009}.
In this scenario, star formation would be regulated by dying star  
ejections and central black hole heating, which in most massive  halos ablate and absorb the infalling cold phase.
These feedback mechanisms have not been explicitly taken
into account in the simulations we present here,
but they are unlikely to suppress cold gas infall
through filaments {\it towards} the centres of the attraction basins
of proto-MGOs at high $z$,  because this has to do
with a singularity formation. 
They can, and very likely
they do, determine the amount of gas infall {\it inside} the halos or centres of  attraction
basins at high $z$ and regulate   star formation processes 
at smaller scales \citep{Slyz:2005}.
This is mimicked in a sense by the star formation parametrisation we use in this paper.
Note, however that the feedback mechanisms or even their 
direction are  still not well known. Indeed,  some authors propound an enhancement of the star formation efficiency
due to AGN feedback, implying suprasolar $\alpha/<$Fe$>$
in massive galaxies \citep{Pipino:2009}, while others find that AGN activity could have the opposite effect
\citep{Kawata:2005}.

To sum up, gravo-hydrodynamical simulations have
allowed us to advance our understanding of massive galaxy formation
in connection with the global cosmological model,
and more particularly of how baryonic matter exchanges at different important scales came about.
The simulations are a virtual experiment that mimic galaxy
formation and evolution, in such a way that they allow us to deepen
into  the physical processes causing them. 
In this  paper, we report on MGO mass assembly and stellar formation 
processes in a series of  simulations that follow 
what could be expected
from the adhesion  and singularity dressing
models, when gas processes are allowed for. Moreover,
direct implications of these models nicely explain different 
generic properties of elliptical galaxies and their samples, some of which
could seem paradoxical within a hierarchical scenario.

Let us stress that the assembly patterns we found here	
are just a consequence of the advanced non-linear stage of
gravitational instability evolution, as described by the
{\it adhesion} model	
within a $\Lambda$CDM cosmological model. 
Otherwise, they result from simple physical laws acting on quite
general initial conditions.
This is an important result and, in a sense,  it represents a test
of the {\it concordance cosmological model}
on the scales relevant to galaxy formation (a few  hundred kpc).

\section*{Acknowledgments}

The authors are grateful to the anonymous referee for his/her help  to improve this paper.
We thankfully acknowledge to  J. Naranjo and D. Vicente for the assistence and technical expertise
provided at
the Barcelona Supercomputing Centre, as well as the computer resources provided by BSC/RES (Spain).
RDT is happy to thank L.J. Roy for the very motivating discussions on singular flows at the beginning of this work.
We thank DEISA Extreme Computing Initiative (DECI) for the CPU time allowed to GALFOBS project.
The Centro de Computaci\'on Cientif\'ica (UAM, Spain) has also provided computing facilities.
This work was partially supported by the DGES (Spain) through the
grants
%s AYA2006-15492-C03-01, AYA2006-15492-C03-02 and
AYA2009-12792-C03-02 and AYA2009-12792-C03-03 from the PNAyA, as well as by the regional Madrid V PRICIT program
through the ASTROMADRID network (CAM S2009/ESP-1496).
JO was supported by the ''Supercomputaci\'on y e-Ciencia'' Consolider-Ingenio CSD2007-0050 project.

\label{lastpage}


\begin{thebibliography}{99}

\bibitem[\protect\citeauthoryear{Arag\'on-Calvo, van de Weygaert \& Jones}{2010}]{Aragon:2010} Arag{\'o}n-Calvo, M.~A., van de Weygaert, R.,
\& Jones, B.~J.~T.\ 2010, \mnras, 408, 2163 

\bibitem[\protect\citeauthoryear{Aretxaga, Kunth \& M\'ujica}{2001}]{Aretx:2001} Aretxaga I., Kunth D.,  M\'ujica, R., 2001, eds, Advanced Lectures on the Starburst-AGN Connection, World Scientific 

\bibitem[\protect\citeauthoryear{Bell et al.}{2004}]{Bell:2004} Bell E.F. et al., 2004, ApJ, 608, 752

\bibitem[\protect\citeauthoryear{Bell et al.}{2006}]{Bell:2005} Bell E.F. et al., 2006, ApJ, 640, 241

\bibitem[\protect\citeauthoryear{Bernardi et al.}{2003a}]{Bernardi:2003a} Bernardi M. et al.,  
 2003a, \aj, 125, 1817 

\bibitem[\protect\citeauthoryear{Bernardi et al.}{2003b}]{Bernardi:2003b} Bernardi M. et al.,  
2003b, \aj, 125, 1849 

\bibitem[\protect\citeauthoryear{Bernardi et al.}{2003c}]{Bernardi:2003c} Bernardi M. et al., 2003c, AJ, 125, 1866

\bibitem[\protect\citeauthoryear{Bernardi et al.}{2003d}]{Bernardi:2003d} Bernardi M. et al., 2003d, AJ, 125, 1882


\bibitem[\protect\citeauthoryear{Beuing et al.}{1999}]{Beuing:1999} Beuing J., Dobereiner S., Bohringer H., Bender R., 1999, \mnras, 302, 209

\bibitem[\protect\citeauthoryear{Binney}{1977}]{Binney:1977} Binney J., 1977, \apj, 215, 
483 

\bibitem[\protect\citeauthoryear{Binney 
\& Tremaine}{2008}]{Binney:2008} Binney J.,  Tremaine S., 2008, Galactic Dynamics: Second Edition,  Princeton, NJ, Princeton University Press, 2008, 361 p.,  



\bibitem[\protect\citeauthoryear{Binney}{2004}]{Binney:2004} Binney J., 2004, MNRAS, 347, 1093

\bibitem[\protect\citeauthoryear{Bridge, Carlberg \& Sullivan}{2010}]{Bridge:2010}    Bridge  C.R., Carlberg, R.G., Sullivan M., 2010, ApJ, 709, 1067

\bibitem[\protect\citeauthoryear{Buchert \& Dom\'{\i}nguez}{1998}]{Buchert:1998} Buchert T., Dom\'{\i}nguez A., 1998, A\&A, 335, 395

\bibitem[\protect\citeauthoryear{Buchert \& Dom\'{\i}nguez}{2005}]{Buchert:2005} Buchert T., Dom\'{\i}nguez A., 2005, A\&A, 438,443 

\bibitem[\protect\citeauthoryear{Buchert, Dom\'{\i}nguez \& P\'erez-Mercader}{1999}]{Buchert:1999} Buchert T., Dom\'{\i}nguez A., P\'erez-Mercader J., 1999, A\&A, 349, 343 

\bibitem[\protect\citeauthoryear{Bundy, Ellis \& Conselice}{2005}]{Bundy:2005} Bundy K., Ellis R.S.,  Conselice C.J., 2005, ApJ, 625, 621 

\bibitem[\protect\citeauthoryear{Bundy et al.}{2009}]{Bundy:2009} Bundy K., Fukugita M., Ellis R.S., Targett T.A., Belli S., Kodama T., 2009, ApJ, 697, 1369 

\bibitem[\protect\citeauthoryear{Burgers}{1948}]{Burgers:1948} Burgers J.M., 1948, The Nonlinear Diffusion Equation: Asymptotic Solutions and Statistical Problem,
Proc. Konink. Nederl. Akad. Wetensch. {\bf 1}, 171--199

\bibitem[\protect\citeauthoryear{Burgers}{1974}]{Burgers:1974} Burgers J.M., 1974, Reidel, Dordrecht


\bibitem[\protect\citeauthoryear{Cimatti et al.}{2002}]{Cimatti:2002} Cimatti A. et al., 2002, A\&A, 381, L68 

\bibitem[\protect\citeauthoryear{Cimatti et al.}{2004}]{Cimatti:2004} Cimatti A. et al., 2004, Nature, 430, 184

\bibitem[\protect\citeauthoryear{Clemens et al.}{2009}]{Clemens:2009} Clemens M.~S., 
Bressan A., Nikolic B., Rampazzo, R., 2009, \mnras, 392, L35 



\bibitem[\protect\citeauthoryear{Conselice, Blackburne \& Papovich}{2005}]{Conselice:2005} Conselice C.J., Blackburne J.A., Papovich, C., 2005, ApJ, 620, 564

\bibitem[\protect\citeauthoryear{Conselice et al.}{2003}]{Conselice:2003} Conselice C. J., Bershady M. A., Dickinson M., Papovich C., \ 2003, AJ, 126, 1183

\bibitem[\protect\citeauthoryear{Conselice et al.}{2009}]{Conselice:2009} Conselice C.~J., 
Yang C., Bluck A.~F.~L., \ 2009, \mnras, 394, 1956 


\bibitem[\protect\citeauthoryear{Cook, Lapi \& Granato}{2009}]{Cook:2009} Cook M., Lapi A.,  Granato G. L., 2009, \mnras, 397, 534


\bibitem[\protect\citeauthoryear{Cowie 
\& Binney}{1977}]{Cowie:1977} Cowie L.~L.,  Binney J., 1977, \apj, 215, 723 


\bibitem[\protect\citeauthoryear{Cowie et al.}{1996}]{Cowie:1996} Cowie L.~L., Songaila 
A., Hu E.~M.,  Cohen J.~G., 1996, \aj, 112, 839 

\bibitem[\protect\citeauthoryear{Diehl \& Statler}{2005}]{Diehl:2005} Diehl S., Statler T.S., 2005, ApJ, 633L, 21

\bibitem[\protect\citeauthoryear{Dom\'inguez}{2000}]{Dominguez:2000} Dom\'inguez A., 2000, Phys. Rev. D, 62, 103501
 


\bibitem[\protect\citeauthoryear{Dom\'{\i}nguez-Tenreiro, S\'aiz \& Serna}{2004}]{DTal:2004} Dom\'{\i}nguez-Tenreiro R., S\'aiz A.,  Serna A., 2004, ApJ, 611L, 5


\bibitem[\protect\citeauthoryear{Dom\'{\i}nguez-Tenreiro et al.}{2006}]{DTal:2006} Dom\'{\i}nguez-Tenreiro R., O\~norbe J., S\'aiz A., Artal H.,  Serna, A., \ 2006, \apj, 636L, 77


\bibitem[\protect\citeauthoryear{Dom\'{\i}nguez-Tenreiro et al.}{2008}]{DTal:2008} Dom\'{\i}nguez-Tenreiro R., O\~norbe J., Serna A., Gonz\'alez-Garc\'{\i}a A. C., \ 2008, ASPC, 390, 468


\bibitem[\protect\citeauthoryear{Drory et al.}{2004}]{Drory:2004} Drory N., Bender R., Feulner G., Hopp U., Maraston C., Snigula J., Hill G.J., 2004, ApJ, 608, 742

\bibitem[\protect\citeauthoryear{Dunkley et al.}{2009}]{Dunkley:2009} Dunkley J. et al., 
2009, \apjs, 180, 306 

\bibitem[\protect\citeauthoryear{Eggen, Lynden-Bell \& Sandage}{1962}]{Eggen:1962} Eggen O.~J., 
Lynden-Bell D., Sandage A.~R., 1962, \apj, 136, 748

\bibitem[\protect\citeauthoryear{Evrard et al.}{1990}]{Evrard:1990} Evrard  A.~E., Silk  J., 
 Szalay A.~S., 1990, \apj, 365, 13 


\bibitem[\protect\citeauthoryear{Faber \& Jackson}{1976}]{Faber:1976}  Faber S. M., Jackson R. E., 1976, \apj, 204, 668

\bibitem[\protect\citeauthoryear{Faber et al.}{2007}]{Faber:2007} Faber S.~M. et al., 
2007, \apj, 665, 265 


\bibitem[\protect\citeauthoryear{Fabian 
\& Nulsen}{1977}]{Fabian:1977} Fabian A.~C.,  Nulsen P.~E.~J., 1977, \mnras, 180, 479 

\bibitem[\protect\citeauthoryear{Falomo et al.}{2008}]{Falomo:2008} Falomo R., Traves A., Kotilainen J., Scarpa R., 2008, ApJ, 673, 694 

\bibitem[\protect\citeauthoryear{Fardal et al.}{2001}]{Fardal:2001} Fardal M.~A., Katz N., 
Gardner J.~P., Hernquist L., Weinberg D.~H., 
 Dav{\'e} R., 2001, \apj, 562, 605 


\bibitem[\protect\citeauthoryear{Ferrarese \& Merritt}{2000}]{Ferrarese:2000} Ferrarese L., Merritt D., 2000, ApJ, 539L, 9

\bibitem[\protect\citeauthoryear{Ferrarese \& Ford}{2005}]{Ferrarese:2005} Ferrarese L., Ford H., 2005, Space Science Reviews, 116, 523

\bibitem[\protect\citeauthoryear{Fontana et al.}{2004}]{Fontana:2004} Fontana A. et al., 2004, A\&A, 424, 23 


\bibitem[\protect\citeauthoryear{Gallazzi et al.}{2006}]{Gallazzi:2006} Gallazzi A., Charlot 
S., Brinchmann J.,  White S.~D.~M., 2006, \mnras, 370, 1106 

\bibitem[\protect\citeauthoryear{Gebhardt et al.}{2000}]{Gebhardt:2000} Gebhardt, K. et al., 2000, ApJ, 539L, 13


\bibitem[\protect\citeauthoryear{Gonz\'alez-Garc\'{\i}a {\&} van Albada}{2003}]{GonzalezGarcia:2003} Gonz\'alez-Garc\'{\i}a A.C.,  van Albada T.S., 2003, MNRAS, 342, 36

\bibitem[\protect\citeauthoryear{Gonz{\'a}lez-Garc{\'{\i}}a et 
al.}{2009}]{GonzalezGarcia:2009} Gonz{\'a}lez-Garc{\'{\i}}a A.~C., O{\~n}orbe J., Dom{\'{\i}}nguez-Tenreiro R.,  G{\'o}mez-Flechoso M.~{\'A}., 2009, \aap, 497, 35 

\bibitem[\protect\citeauthoryear{Governato et al.}{2004}]{Governato:2004} Governato F. et al., 2004, ApJ, 607, 688

\bibitem[Granato et al.(2004)]{Granato:2004} Granato, G.~L., De 
Zotti, G., Silva, L., Bressan, A., \& Danese, L.\ 2004, \apj, 600, 580 


\bibitem[\protect\citeauthoryear{Gurbatov 
\& Saichev}{1984}]{Gurbatov:1984} Gurbatov S.~N., Saichev A.~I., 1984, Radiofizika, 27, 456 

\bibitem[\protect\citeauthoryear{Gurbatov et al.}{1989}]{Gurbatov:1989} Gurbatov S.~N., 
Saichev A.~I.,  Shandarin S.~F., 1989, \mnras, 236, 385 

\bibitem[\protect\citeauthoryear{Gurbatov et al.}{1991}]{Gurbatov:1991} Gurbatov S.N., Malakhov A., Saichev A.I., 1991,
"Nonlinear Random Waves and Turbulence in Nondispersive Media", Manchester University Press, Manchester


\bibitem[\protect\citeauthoryear{Hennebelle \& Chabrier}{2010}]{Henne:2009} Hennebelle P.,  Chabrier G., 2009, \apj, 702, 1428 

\bibitem[\protect\citeauthoryear{Hyde 
\& Bernardi}{2009}]{Hyde:2009} Hyde J.~B.,  Bernardi M., 2009, \mnras, 396, 1171 



\bibitem[\protect\citeauthoryear{Jim\'enez et al.}{2007}]{Jimenez:2007} Jim\'enez R., Bernardi 
M., Haiman Z., Panter B.,  Heavens  A.~F., 2007, \apj, 669, 947 


\bibitem[\protect\citeauthoryear{Jones, van de Weygaert \& Arag\'on-Calvo}{2010}]{Jones:2010} Jones, B.~J.~T., van de
Weygaert, R., \& Arag{\'o}n-Calvo, M.~A.\ 2010, \mnras, 408, 897 

\bibitem[\protect\citeauthoryear{Kartaltepe et al.}{2007}]{Kartaltepe:2007} Kartaltepe J.~S.
et al.,\ 2007, \apjs, 172, 320 

\bibitem[\protect\citeauthoryear{Katz}{1992}]{Katz:1992} Katz N., 1992, \apj, 391, 502


\bibitem[\protect\citeauthoryear{Katz et al.}{2003}]{Katz:2003} Katz N., Kere\v s D., Dave R.,
Weinberg, D.~H., 2003, The IGM/Galaxy Connection.~The Distribution of Baryons at z=0, 281, 185

\bibitem[\protect\citeauthoryear{Kauffmann et~al.}{2003{\natexlab{a}}}]{Kauffmann:2003a}
{Kauffmann} G. et al.,  2003{\natexlab{a}}, \mnras, 341, 33


\bibitem[\protect\citeauthoryear{Kauffmann et~al.}{2003{\natexlab{b}}}]{Kauffmann:2003b}
{Kauffmann} G. et al., 2003{\natexlab{b}},
  \mnras, 341, 54

\bibitem[\protect\citeauthoryear{Kauffmann et~al.}{2003{\natexlab{c}}}]{Kauff:2003} Kauffmann G. et al., 2003{\natexlab{c}}, \mnras, 346, 1055 


\bibitem[\protect\citeauthoryear{Kay et al.}{2000}]{Kay:2000} Kay S.~T., Pearce F.~R., 
Jenkins A., Frenk C.~S., White S.~D.~M., Thomas P.~A.,  Couchman H.~M.~P., 2000, \mnras, 316, 374 


\bibitem[\protect\citeauthoryear{Kawata \& Gibson}{2005}]{Kawata:2005} Kawata D., Gibson B., 2005, ApJ, 358L, 16

\bibitem[\protect\citeauthoryear{Kere{\v s} et al.}{2009}]{Keres:2009} Kere{\v s} D., 
Katz N., Fardal M., Dav{\'e} R., Weinberg D.~H., 2009, \mnras, 395, 160 



\bibitem[\protect\citeauthoryear{Knebe, Dom{\'\i}nguez \& Dom{\'\i}nguez-Tenreiro}{2006}]{Knebe:2006} Knebe A., Dom{\'\i}nguez A., \& Dom{\'\i}nguez-Tenreiro R., 2006, \mnras, 371, 1959

\bibitem[\protect\citeauthoryear{Kofman, Pogosyan \& Shandarin}{1990}]{Kofman:1990}  Kofman L., Pogosyan D., Shandarin S., 1990, \mnras, 242, 200

\bibitem[\protect\citeauthoryear{Larson}{1974}]{Larson:1974} Larson R.~B., 1974, \mnras, 
166, 585 

\bibitem[\protect\citeauthoryear{Le F\`evre et al.}{2000}]{LeFreve:2000} Le F\`evre O. et al., 2000, MNRAS, 311, 565

\bibitem[\protect\citeauthoryear{Lin et al.}{2008}]{Lin:2008} Lin L. et al., 2008, ApJL, 681, 232L

\bibitem[\protect\citeauthoryear{Lotz et al.}{2008}]{Lotz:2008} Lotz J. et al., 2008, ApJL, 672, 177L

\bibitem[\protect\citeauthoryear{Magorrian et al.}{1998}]{Magorrian:1998} Magorrian J. et al., 1998, AJ, 115, 2285

\bibitem[\protect\citeauthoryear{Maraston et 
al.}{2003}]{Maraston:2003} Maraston C., Greggio L., Renzini A., Ortolani S., Saglia R.~P., Puzia T.~H., Kissler-Patig M., 2003, \aap, 400, 823

\bibitem[\protect\citeauthoryear{Mart{\'{\i}}nez-Serrano}{2010}]{TesisFran} Mart{\'{\i}}nez-Serrano F. J., 2010, PhD thesis, Universidad Aut\'onoma de Madrid, http://symmetry.ft.uam.es/fran/PhD\_Francisco\_Martinez.pdf

\bibitem[\protect\citeauthoryear{Mart{\'{\i}}nez-Serrano et al.}{2009}]{MartinezSerrano:2009} 
Mart{\'{\i}}nez-Serrano F.~J., Serna A., Dom{\'e}nech-Moral M., 
Dom{\'{\i}}nguez-Tenreiro R., \ 2009, ApJ, 705L, 133 

\bibitem[\protect\citeauthoryear{Mathews 
\& Brighenti}{2003}]{Mathews:2003} Mathews W.~G.,  Brighenti F., \ 2003, \araa, 41, 191 

\bibitem[\protect\citeauthoryear{Matteucci}{2003}]{Matteucci:2003} Matteucci F., 2003, Ap\&SS, 284, 539

\bibitem[\protect\citeauthoryear{McIntosh et al.}{2005}]{McIntosh:2005} McIntosh D.H. et al., 2005, ApJ, 632, 191

\bibitem[\protect\citeauthoryear{Menanteau et al.}{2004}]{Menanteau:2004} Menanteau F. et al., 2004, ApJ, 612, 202

\bibitem[\protect\citeauthoryear{Mobasher et al.}{2005}]{Mobasher:2005} Mobasher B. et al., 2005, ApJ, 635, 832

\bibitem[\protect\citeauthoryear{Mobasher \& Wiklind}{2010}]{Mobasher:2010} Mobasher B., Wiklind T., 2010, in  The Impact of HST on European Astronomy, Astrophysics and Space Science Proceedings, F. Duccio Macchetto ed., Springer, Netherlands 

\bibitem[\protect\citeauthoryear{Navarro \& White}{1994}]{Navarro:1994} Navarro J.F., White S.D.M., 1994, MNRAS, 267, 401

\bibitem[\protect\citeauthoryear{Neichel et al.}{2008}]{Neichel:2008} Neichel B. et al., 2008, A\&A, 484, 159


\bibitem[\protect\citeauthoryear{O\~norbe}{2009}]{TesisJose} O\~norbe J., \ 2009, PhD thesis,  Universidad Aut\'onoma de Madrid, http://jonorbe.ps.uci.edu/onorbe/doc/thesis-jonorbe.pdf

\bibitem[\protect\citeauthoryear{O\~norbe et al.}{2005}]{Onorbe:2005} O\~norbe J., Dom\'{\i}nguez-Tenreiro  R., S\'aiz  A., Serna  A., Artal, H., 2005, ApJ, 632L, 570 

\bibitem[\protect\citeauthoryear{O{\~n}orbe et al.}{2006}]{Onorbe:2006} O{\~n}orbe J., 
Dom{\'{\i}}nguez-Tenreiro R., S{\'a}iz A., Artal H., 
Serna A., \ 2006, \mnras, 373, 503 

\bibitem[\protect\citeauthoryear{O{\~n}orbe et al.}{2007}]{Onorbe:2007} O{\~n}orbe, J., 
Dom{\'{\i}}nguez-Tenreiro R., S{\'a}iz A., 
Serna A., \ 2007, \mnras, 376, 39 

\bibitem[\protect\citeauthoryear{O{\~n}orbe et al.}{2011}]{Onorbe:2011} O{\~n}orbe, J., 
Mart\'inez-Serrano F.J., Dom{\'{\i}}nguez-Tenreiro R., Knebe A., Serna A., \ 2011, ApJL, in press. (http://arxiv.org/abs/1103.4214).

\bibitem[\protect\citeauthoryear{Oser et al.}{2010}]{Oser:2010} Oser, L., Ostriker, J.~P.,
Naab, T., Johansson, P.~H., \& Burkert, A.\ 2010, \apj, 725, 2312 

\bibitem[\protect\citeauthoryear{Padmanabhan}{1993}]{Padma:1993} Padmanabhan T., \ 1993, in Structure Formation in the Universe, Ch. 8, Cambridge University Press

\bibitem[\protect\citeauthoryear{Patton et al.}{2002}]{Patton:2002} Patton D.R et al., 2002, ApJ, 565, 208

\bibitem[\protect\citeauthoryear{Pipino, Silk \& Matteucci}{2009}]{Pipino:2009} Pipino A., Silk J., Matteucci F., 2009, \mnras, 392, 475


\bibitem[\protect\citeauthoryear{Power \& Knebe}{2006}]{Power:2006} Power C., Knebe A.,\ 2006, \mnras, 370, 691

\bibitem[\protect\citeauthoryear{Press \& Schechter}{1974}]{Press:1974} Press W. H.,  Schechter P., \ 1974, \apj, 187, 425 


\bibitem[\protect\citeauthoryear{Rees 
\& Ostriker}{1977}]{Rees:1977} Rees M.~J., Ostriker J.~P.\ 1977, \mnras, 179, 541 


\bibitem[\protect\citeauthoryear{S{\'a}iz et al.}{2001}]{Saiz:2001} S{\'a}iz A., 
Dom{\'{\i}}nguez-Tenreiro R., Tissera P.~B., 
Courteau S., \ 2001, \mnras, 325, 119


\bibitem[\protect\citeauthoryear{S{\'a}iz, Dom{\'{\i}}nguez-Tenreiro
\& Serna}{2004}]{Saiz:2004} S{\'a}iz A., Dom{\'{\i}}nguez-Tenreiro R., Serna A.,\ 2004, ApJ, 601L, 131


\bibitem[\protect\citeauthoryear{Salvador-Sol\'e, Manrique \& Solanes}{2005}]{Salvador:2005} Salvador-Sol\'e, E.,  Manrique, A., Solanes, J.M., \ 2005, MNRAS, 358, 901

\bibitem[\protect\citeauthoryear{S{\'a}nchez-Bl{\'a}zquez et al.}{2006}]{Sanchezblazquez:2006} 
S{\'a}nchez-Bl{\'a}zquez, P., et al.\ 2006, \mnras, 371, 703 

\bibitem[\protect\citeauthoryear{Serna, Dom{\'{\i}}nguez-Tenreiro, \& S\'aiz}{2003}]{Serna:2003} Serna A., Dom{\'{\i}}nguez-Tenreiro R., S\'aiz, A., 2003, ApJ, 597, 878

\bibitem[\protect\citeauthoryear{Shandarin 
\& Zeldovich}{1989}]{Shandarin:1989} Shandarin S.~F.,  Zeldovich Y.~B.\ ,1989, Reviews of Modern Physics, 61, 185 

\bibitem[\protect\citeauthoryear{Silk \& Nusser}{2010}]{Silk:2010} Silk, J., \& Nusser, A.\ 2010, \apj, 725, 556 

\bibitem[\protect\citeauthoryear{Slyz et al.}{2005}]{Slyz:2005} Slyz A.~D., Devriendt 
J.~E.~G., Bryan G., Silk J.,\ 2005, \mnras, 356, 737 


\bibitem[\protect\citeauthoryear{Sparks et al.}{1989}]{Sparks:1989} Sparks W.~B., 
Macchetto F.,  Golombek D.,\ 1989, \apj, 345, 153 

\bibitem[\protect\citeauthoryear{Stanford et al.}{2004}]{Stanford:2004} Stanford, S.~A.,
Dickinson, M., Postman, M., Ferguson, H.~C., Lucas, R.~A., Conselice,
C.~J., Budav{\'a}ri, T., \& Somerville, R.\ 2004, \aj, 127, 131 


\bibitem[\protect\citeauthoryear{Thacker \& Couchman}{2000}]{Thacker:2000} Thacker R.J.,  Couchman, H.M.P., 2000, ApJ, 545, 728

\bibitem[\protect\citeauthoryear{Thomas, Greggio \& Bender}{1999}]{Thomas:1999} Thomas D.,  Greggio L.,  Bender R., 1999, MNRAS, 302, 537


\bibitem[\protect\citeauthoryear{Thomas et al.}{2005}]{Thomas:2005} Thomas D.,  Maraston C., Bender R., Mendes de Oliveira, C., 2005, ApJ, 621, 673

\bibitem[\protect\citeauthoryear{Tinsley}{1972}]{Tinsley:1972} Tinsley B.~M.,\ 1972, \apj, 
178, 319 


\bibitem[\protect\citeauthoryear{Tissera, Lambas \& Abadi}{1997}]{Tissera:1997} Tissera P.B., Lambas D.G.,  Abadi M.C., 1997, \mnras, 286, 384

\bibitem[\protect\citeauthoryear{Toomre}{1977}]{Toomre:1977} Toomre A., 1977, in The Evolution of Galaxies and Stellar Populations, eds.\ B. Tinsley \& R. Larson (New Have, CN: Yale Univ.\ Press)


\bibitem[\protect\citeauthoryear{Treu \& Koopmans}{2004}]{Treu:2004} Treu T.,   Koopmans, L.V.E., 2004, ApJ, 611, 739

\bibitem[\protect\citeauthoryear{Trujillo et al.}{2004}]{Trujillo:2004} Trujillo I. et al., 2004, ApJ, 604, 621


\bibitem[\protect\citeauthoryear{van der Wel et al.}{2004}]{vanderWel:2004} van der Wel A., Franx M., van Dokkum P.G.,  Rix H.-V., 2004, ApJ, 601L, L5


\bibitem[\protect\citeauthoryear{van Dokkum \& Ellis}{2003}]{vanDokkum:2003} van Dokkum P.G., Ellis R.S., 2003, ApJ, 592, L53

\bibitem[\protect\citeauthoryear{Vergassola et al.}{1994}]{Vergassola:1994} Vergassola ~M., Dubrulle ~B., Frisch ~U., Noullez,~A., 1994, A\&A, 289, 325

\bibitem[\protect\citeauthoryear{Wechsler et al.}{2002}]{Wechsler:2002}
Wechsler R.H., Bullock J.S., Primack J.R., Kravtsov A.V., Dekel A., 2002, ApJ, 568, 52

\bibitem[\protect\citeauthoryear{Weinberg \& Gunn}{1990}]{Weinberg:1990}  Weinberg D. H., Gunn J. E., 1990, \mnras, 247, 260 

\bibitem[\protect\citeauthoryear{White 
\& Rees}{1978}]{White:1978} White S.~D.~M.,  Rees M.~J., \ 1978, \mnras, 183, 341 

\bibitem[\protect\citeauthoryear{Wiklind et al.}{2008}]{Wiklind:07} Wiklind T., Dickinson M., Ferguson H.~C., Giavalisco M., Mobasher B., Gorgin N.~A., Panagia N., \ 2008, ApJ, 676, 781


\bibitem[\protect\citeauthoryear{York et~al.}{2000}]{York:2000}
{York} D.~G. et al., 2000, \aj, 120, 1579

\bibitem[\protect\citeauthoryear{Zeldovich}{1970}]{Zeldovich:1970} Zeldovich Y.~B.,\ 1970, \aap, 5, 84 


\bibitem[\protect\citeauthoryear{Zhao et al.}{2003}]{Zhao:2003}
Zhao D.H., Mo H.J., Jing Y.P.,  Borner G., 2003, MNRAS, 339, 12


\bibitem[\protect\citeauthoryear{Zheng et al.}{2005}]{Zheng:2005}
Zheng, X.~Z., Hammer, F., Flores, H., Ass{\'e}mat, F., \& Rawat, A.\ 2005, \aap, 435, 507


\end{thebibliography}
\end{document}